\documentclass[lettersize,journal]{IEEEtran}
\usepackage{amsthm}    
\usepackage{amsmath}   
\usepackage{amsfonts}
\usepackage{amssymb}   
\usepackage{algorithmic}
\usepackage{algorithm}
\usepackage{array}
\usepackage[caption=false,font=normalsize,labelfont=sf,textfont=sf]{subfig}
\usepackage{textcomp}
\usepackage{stfloats}
\usepackage{soul}
\usepackage{xcolor}
\sethlcolor{yellow}
\usepackage{url}
\usepackage{verbatim}
\usepackage{comment}
\usepackage{siunitx}
\usepackage{graphicx}
\usepackage{cite}
\usepackage[hidelinks]{hyperref}
\usepackage{orcidlink}
\usepackage{mathtools}
\usepackage{tikz}

\soulregister\cite7
\soulregister\ref7
\soulregister\eqref7
\soulregister\pageref7

\theoremstyle{plain}
\newtheorem{lemma}{Lemma}[section]        
\newtheorem{proposition}{Proposition}[section]
\newtheorem{definition}{Definition}[section]

\theoremstyle{remark}
\newtheorem{remark}{Remark}[section]

\begin{document}

\title{Fault Detection under Uncertainty: Evaluation of Observer Designs on a Wind-Farm Benchmark}
\author{\'{A}lvaro Mart\'{\i}n G\'{o}mez\orcidlink{0009-0005-6122-626X},
Szymon Greś\orcidlink{0000-0002-6566-2713},
Sina Hassani\orcidlink{0000-0003-0459-9738},
Yerai Peña-Sánchez\orcidlink{0000-0001-8614-0837},
Rafał Wiśniewski\orcidlink{0000-0001-6719-8427},
and Steven X.~Ding\orcidlink{0000-0002-5149-5918}%
\thanks{\'{A}.~M. G\'{o}mez, S. Greś, S. Hassani, and R. Wiśniewski are with the Department of Electronic Systems, Aalborg University, Aalborg, Denmark (e-mail: \mbox{almago@es.aau.dk}; \mbox{sgres@es.aau.dk}; \mbox{sinah@es.aau.dk}; \mbox{raf@es.aau.dk}).}%
\thanks{Y. P.~Sánchez is with the Polytechnic School of Mechanical and Industrial Production Engineering, Mondragón University, Mondragón, Spain (e-mail: \mbox{ypena@mondragon.edu}).}%
\thanks{S. X.~Ding is with the Department of Automatic Control and Complex Systems, University of Duisburg--Essen, Duisburg, Germany (e-mail: steven.ding@uni-due.de).}%
\thanks{This work was supported by the BeWind project, funded by Innovation Fund Denmark.}%
}



\maketitle

\begin{abstract}
This paper presents a robust, model–based anomaly detection framework for an offshore wind farm, implemented on a standardized Simulink benchmark for reproducible evaluation. Two observer–based residual generators are compared: an orthogonal–projection filter based on the stable kernel representation of the system, and an $\mathcal H_i/\mathcal H_\infty$ trade-off filter (sensitivity to faults vs. robustness to disturbances), typically referred to as the unified solution. Parametric uncertainty in the aerodynamic model is propagated through the Riccati solution, gains, and post–filters up to second order. The first-order covariance approximation is used for statistical change detection, while second-order residual bounds are used to assess the effect of nonlinear uncertainty propagation. The two filter designs induce distinct residual behaviours and uncertainty bands; this relationship is analysed.
In the detection stage, the residual distribution under parametric uncertainty is explicitly analysed. A first–order approximation leads to a Gaussian characterization and a generalized likelihood ratio test, while second–order terms lead to a generalised $\chi^2$ distribution and motivate alternative, distribution-free thresholding strategies. Overall, the framework provides a benchmarked assessment of observer-based fault detection performance under explicitly quantified model uncertainty.
\end{abstract}


\section{Introduction}

\IEEEPARstart{T}{he} reliability and availability of modern wind turbines (WTs) depend strongly on effective fault detection and isolation (FDI) strategies in their aerodynamic and mechanical subsystems. These components, comprising the blades, pitch and yaw systems, drive-train, generator, and tower, are subject to a wide variety of failure modes such as blade icing and erosion, pitch misalignment, actuator or sensor degradation, drivetrain bearing wear, and tower–nacelle structural defects. The growing complexity of wind energy conversion systems, particularly for offshore and floating configurations, has intensified the need for integrated diagnostic frameworks that can operate under strong dynamic couplings and environmental uncertainties.

Model-based fault diagnosis remains a foundational approach in the field. Observer- and parity-equation-based methods exploit analytical redundancy to generate residuals sensitive to faults but robust to disturbances. Classical designs include Kalman and unknown-input observers (UIO), parity relations, and robust $\mathcal{H}_{\infty}$ filters \cite{odgaard2012fault,Pisu2011_IFAC,Ozdemir2011_IFAC}. These frameworks are valued for their low data dependency, interpretability, and clear statistical structure for decision-making via generalized likelihood ratio (GLR) or CUSUM tests\cite{basseville1993detection}.

Recent model-based research has increasingly focused on constraint-aware observers and predictive residual generators that can exploit high-resolution time-series data (e.g., 10–100 ms sampling) to capture the transient and nonlinear aero-servo-elastic couplings characteristic of modern turbines \cite{verma2022wind}. Mixed Bayesian/set-membership formulations, robust and sliding-mode observers, and predictive/FTC-oriented estimators enhance robustness to uncertainty while improving sensitivity to incipient pitch and drivetrain faults \cite{FernandezCanti2015_CEP,bououden2023observer,borja2023actuator,lan2018fault}. For floating turbines, dedicated observer extensions explicitly incorporate platform-motion states, enabling reliable fault isolation under the strong coupled interactions between aerodynamics, hydrodynamics, and the turbine’s control system \cite{cho2018model,ammerman2024experimental,wang2024fault}. These developments reflect a broader trend toward unified estimation–control structures where model-based FDI operates coherently with supervisory and fault-tolerant control layers.


Complementing model-based strategies, recent data-driven and hybrid learning methods increasingly use high-frequency operational data to model normal turbine behaviour and detect early deviations in pitch, drive-train, and generator subsystems. Studies show that autoencoders, convolutional/recurrent networks, and attention-based models can capture transient fault signatures in multivariate time series, often outperforming traditional SCADA-based diagnostics for subtle or early-stage failures \cite{elshenawy2024fault,xiang2021fault,castellani2024wind}. Physics-informed and hybrid approaches further improve robustness by embedding simplified aeroelastic or drive-train structure, reducing the need for large labelled datasets \cite{nogueira2025wind}. At the same time, transfer-learning and domain-adaptation methods help extend models across different turbines and sites when appropriate feature alignment is applied \cite{Vervlimmeren2025_WESpreprint}. Overall, these developments show that high-resolution, data-driven FDI is becoming a practical complement to observer-based techniques, particularly when fast dynamics and nonlinear interactions are diagnostically important.

Despite remarkable progress, several open challenges remain:
\begin{itemize}
  \item \emph{Generalization and domain shift.}
  Data-driven models often fail to generalize across sites and turbine types without adaptation \cite{Vervlimmeren2025_WESpreprint}.
  \item \emph{Benchmarking and data access.}
  The lack of standardized, multi-site open datasets limits fair comparison of algorithms \cite{Stetco2019_RenewEnergy}.
  \item \emph{Explainability and operator trust.}
  Industrial adoption increasingly requires interpretable decision metrics and fault-symptom mapping \cite{facchini2025wind}.
  \item \emph{Handling model uncertainty.}
  Many anomaly detection frameworks still struggle to account for model and operating-condition uncertainty, limiting their robustness and reliability\cite{tabatabaeipour2012fault}.
\end{itemize}

Overall, the field is moving toward multi-modal fusion, physics-informed learning, and standardized evaluation pipelines, all of which are aligned with the requirements of modern and floating wind turbines.

This paper presents a unified framework for observer-based fault detection that explicitly accounts for parametric uncertainty in wind-turbine models. The contributions are threefold. First, uncertainty is propagated through the full observer design up to second order, enabling a rigorous statistical characterization of the residuals under uncertainty; in particular, the paper derives and provides analytical expressions for the Jacobians and Hessians required for first- and second-order propagation. Second, the paper addresses a gap in the literature by providing a systematic comparison between two observer designs—an orthogonal-projection filter based on the stable kernel representation (SKR) of the system\cite{ding2026alternative}, whose performance has not yet been thoroughly assessed in realistic industrial settings, and the so-called $\mathcal{H}_i/\mathcal{H}_\infty$ trade-off filter (or the "unified solution")\cite{ding2008model}—highlighting how their structures affect residual behaviour, fault sensitivity, and robustness under uncertainty. Third, the first-order covariance approximation is combined with a practical, distribution-free thresholding procedure, while the second-order bounds are used for uncertainty assessment, and the overall approach is validated on the \emph{FOWLTy} wind farm benchmark\cite{pena2024development} under a series of realistic fault scenarios for reproducible evaluation. The focus is on sensor and actuator faults in the blade-pitch and generator-torque systems and on generator-speed and electrical-power sensor faults. Structural, hydrodynamic, wind-speed-sensor and tower-acceleration-sensor faults are outside the evaluation scope. The aim is to study how a low-order wind-turbine model, based on one-dimensional momentum theory for the aerodynamics, can be used for anomaly detection under model uncertainty. Rather than developing a dedicated detector for each subsystem, we use a common detection framework and focus on limiting false alarms. The detection sensitivity achievable with this simplified model is then assessed.

The remainder of this paper is organized as follows. Sec.~\ref{sec:modelling} presents the dynamic modelling of the seven-turbine floating wind-farm benchmark, including aerodynamic, mechanical, and control subsystems. Sec.~\ref{sec:fault_scenarios} describes the considered sensor and actuator fault scenarios used for evaluation. Sec.~\ref{sec:stat} analyses the residual distribution under uncertainty and discusses threshold selection. Sec.~\ref{sec:implementation} outlines the specific implementation details. Sec.~\ref{sec:results} reports and discusses the results, comparing uncertainty behaviour, sensitivity analyses, and detection performance. Sec.~\ref{sec:conclusion} concludes the paper and highlights directions for future work. In addition, Appendix~\ref{appendixA} details the fault-detection filters used in this work, and Appendix~\ref{appendixB} provides the analytical expressions for the uncertainty propagation.

\section{Modelling of a Wind Farm}\label{sec:modelling}

The \emph{FOWLTy} benchmark considers a wind farm composed of seven NREL 5~MW reference floating wind turbines (WTs) and a measuring mast\cite{pena2024development}. Even though the simulation model includes wake effects, they are not considered in the fault-detection algorithm, which operates locally at each WT using only its own measurements. Hydrodynamic effects are also not explicitly considered in order to avoid additional modelling complexity. For turbine $i\in\{1,\dots,7\}$ the effective rotor–plane inflow is modeled as
\begin{equation}
    v_r^i = v_m^i + v_t^i - \dot{x}_t^i,
\end{equation}
where $v_m^i$ is a slowly varying mean component, $v_t^i$ a turbulent fluctuation, and $\dot{x}_t^i$ the fore–aft tower–top velocity, which reduces or increases the apparent inflow when the tower moves toward or away from the wind.

\subsection{Stochastic wind components}
To avoid an unbounded random walk while retaining low–frequency variability, $v_m^i$ follows an Ornstein–Uhlenbeck process
\begin{equation}
    dv_m^i = -\lambda_m v_m^i\,dt + \sigma_m\,dW_m^i,\label{eqn:vm}
\end{equation}
with correlation time $\tau_m=1/\lambda_m$ and Wiener process $W_m^i$. Over horizon $T$, $\mathrm{Var}(v_m^i(T))=\frac{\sigma_m^2}{2\lambda_m}\bigl(1-e^{-2\lambda_m T}\bigr)$. Imposing a design variation of $2~\mathrm{m/s}$ over $T=\SI{600}{s}$ yields $\sigma_m=\sqrt{4\lambda_m/(1-e^{-1200\lambda_m})}$; with $\tau_m=\SI{300}{s}$ we obtain $\sigma_m\approx \SI{0.117}{m/\sqrt{s}}$.

The turbulent component $v_t^i$ is modeled as first-order filtered white noise \cite{knudsen2013simple} with bandwidth matched to the Kaimal spectrum peak at $\omega_p(v_m^i)=\frac{\pi v_m^i}{2L}$\cite{burton2011wind}, where $L=8.1\Lambda_1=\SI{340.2}{m}$ is the longitudinal turbulence length scale and $\Lambda_1=\SI{42}{m}$ is the IEC turbulence scale parameter for hub heights at or above $\SI{60}{m}$ \cite{iec200561400}. The turbulent component satisfies
\begin{equation}
    dv_t^i = -\omega_p(v_m^i)v_t^i\,dt + \omega_p(v_m^i)\sigma_t(v_m^i)\,dW_t^i.\label{eqn:vt}
\end{equation}
The steady-state variance is $\mathrm{Var}(v_t^i)=\frac{\omega_p(v_m^i)\sigma_t(v_m^i)^2}{2}=(t_i v_m^i)^2$ so $\sigma_t(v_m^i)=t_i v_m^i\sqrt{\tfrac{2}{\omega_p(v_m^i)}}$, with turbulence intensity $t_i=\sqrt{\mathrm{Var}(v_t^i)}/v_m^i$ and Wiener process $W_t^i$.

\subsection{Aerodynamics}
Under one–dimensional momentum theory for an ideal rotor, the power and thrust are\cite{hansen2015aerodynamics}
\begin{align}
    P^i &= \tfrac{1}{2}\rho A (v_r^i)^3 C_p(\lambda^i,\beta^i),\\
    F^i &= \tfrac{1}{2}\rho A (v_r^i)^2 C_t(\lambda^i,\beta^i),
    \label{eq:aero_PT}
\end{align}
with air density $\rho$, rotor area $A$, blade pitch $\beta^i$, and tip–speed ratio $\lambda^i=\omega_r^i R/v_r^i$ (rotor speed $\omega_r^i$, radius $R$). The power and thrust coefficients $C_p$ and $C_t$ are typically given in the form of a look-up table as a function of $\beta$ and $\lambda$. In the absence of proprietary lookup tables, we use smooth surrogates\cite{carpintero2020wind}
\begin{align}
    C_p(\lambda,\beta) &= c_1\!\left(c_2\lambda_i^{-1}-c_3\beta-c_4\right)\!e^{-c_5\lambda_i^{-1}},\\
    \lambda_i^{-1} &= \frac{1}{\lambda+0.08\beta}-\frac{0.035}{\beta^3+1},\\
    C_t(\lambda,\beta) &= \sum_{i=0}^2\sum_{j=0}^2 K_{ij}\,\lambda^i\beta^j,
\end{align}
and the aerodynamic torque
\begin{equation}
    T_r^i = \frac{P^i}{\omega_r^i} 
           = \frac{1}{2}\rho A (v_r^i)^3 \frac{C_p(\lambda^i,\beta^i)}{\omega_r^i}.
\end{equation}
Dynamic inflow corrections \cite{knudsen2013simple} are neglected; their effect is dominated here by parametric uncertainty in $C_p,C_t$ for FDI purposes.

\subsection{Drive train, tower, and actuators}
Let $\omega_g^i$ denote the generator-side angular speed of turbine $i$. With a rigid drive train, $\omega_g^i=N\omega_r^i$ and
\begin{equation}
    \dot{\omega}_r^i\,(J_r+N^2J_g) = T_r^i - N T_g^i,
\end{equation}
with rotor and generator inertias $J_r,J_g$, gearbox ratio $N$, and generator torque $T_g^i$. Measured electrical power is $P_{\mathrm{meas}}^i=\eta T_g^i \omega_g^i$ (efficiency $\eta$). Fore–aft tower motion is a mass–spring–damper
\begin{equation}
    \ddot{x}_t^i = \frac{1}{m_{t,\mathrm{eq}}}\bigl(F^i - k_t x_t^i - c_t \dot{x}_t^i\bigr),
\end{equation}
with equivalent mass $m_{t,\mathrm{eq}}$, stiffness $k_t$, damping $c_t$ set from the first natural frequency and damping ratio. Pitch and torque actuators are
\begin{equation}
    \ddot{\beta}^i = \omega_\beta^2(\beta_{\mathrm{ref}}^i-\beta^i) - 2\zeta_\beta\omega_\beta \dot{\beta}^i,
    \qquad
    \dot{T}_g^i = \frac{T_{g,\mathrm{ref}}^i - T_g^i}{\tau_g}.
\end{equation}

\subsection{Baseline control (below- and above-rated operation)}
The baseline controller of the NREL 5~MW reference turbine operates in two distinct regions, depending on the mean wind speed relative to the rated value (approximately $v_{\mathrm{rated}} \simeq 11.4~\mathrm{m/s}$). 
The control objectives are (i) to maximize power capture below rated conditions and (ii) to regulate generator speed and power above rated conditions. The two regimes are:

\begin{itemize}
    \item \textit{Below-rated region}: 
    the turbine operates at a constant blade pitch angle $\beta^i \equiv 0$, and the generator torque is commanded according to the optimal torque control law
    \begin{equation}
        T_{g,\mathrm{ref}}^i = K \, (\omega_r^i)^2,
        \qquad 
        K = \tfrac{1}{2}\frac{\rho R^3 A C_{p,\max}}{\lambda_{\mathrm{opt}}^3},
    \end{equation}
    ensuring operation near the maximum power coefficient $C_{p,\max}$ at the optimal tip-speed ratio $\lambda_{\mathrm{opt}}$.
    
    \item \textit{Above-rated region}: 
    the controller regulates generator speed to its rated value $\omega_{g,\mathrm{rat}}$ by varying the collective pitch and torque commands. 
    The pitch angle is governed by a PI controller acting on the generator speed error
    \begin{equation}
        \beta_{\mathrm{ref}}^i = -K_{p,\beta}\,\dot{\omega}_{g,\mathrm{err}}^i + K_{i,\beta}\,\omega_{g,\mathrm{err}}^i,
    \end{equation}
    with $\omega_{g,\mathrm{err}}^i=\omega_g^i-\omega_{g,\mathrm{rat}}$, while the generator torque command ensures the demanded electrical power
    \begin{equation}
        T_{g,\mathrm{ref}}^i = \frac{P_{\mathrm{dem}}^i}{\eta\,\omega_g^i},
    \end{equation}
    where $\eta$ is the drivetrain efficiency. 
    The demanded power is allocated at farm level as
    \begin{align}
        P_{\mathrm{dem}}^i &= \max\!\left(\hspace{-0.5mm}0,\,\min\!\left(\hspace{-0.5mm}P_{\mathrm{rated}},\,\frac{P_{\mathrm{dem,net}} P_{\mathrm{avail}}^i}{\sum_{j=1}^{7} P_{\mathrm{avail}}^j}\hspace{-0.5mm}\right)\hspace{-0.5mm}\right),\\
        P_{\mathrm{avail}}^i &= \tfrac{1}{2}\rho A (v_r^i)^3 C_p(\lambda^i,\beta^i),
    \end{align}
    with $P_{\mathrm{dem,net}}$ denoting the total power demand from the network operator.
\end{itemize}

\subsection{State-space model}\label{subsec:state_space_model}
For each WT,
\begin{align}
x^i&=\big[\omega_g^i,\; T_g^i,\; \beta_{1,2,3}^i,\; \dot{\beta}_{1,2,3}^i,\; v_m^i,\; v_t^i,\; x_t^i,\; \dot{x}_t^i\big]^\top\label{eqn:x_i},\\
u^i&=\big[T_{g,\mathrm{ref}}^i,\; \beta_{1,2,3,\mathrm{ref}}^i\big]^\top,\label{eqn:u_i}\\
y^i&=\big[\omega_g^i,\; T_g^i,\; \beta_{1,2,3}^i,\; v_r^i,\; \ddot{x}_t^i,\; P_{\mathrm{meas}}^i\big]^\top\label{eqn:y_i}.
\end{align}
Each turbine model is linearized about operating points parametrized by the mean wind speed and then discretized to obtain $(A,B,C,D)$. The full state-space model includes a disturbance vector $d^i$ that stacks two process–noise channels ($v_m^i,v_t^i$) and sensor noises for all measured outputs. The distribution matrix $E_d$ in \eqref{eq:process} assigns unit weights to the diagonal entries corresponding to \((v_m,v_t)\), and the distribution matrix $F_d$ in \eqref{eq:measurement} assigns unit weights to all diagonal entries to account for sensor noise. Furthermore, the fault vector $f$ collects all sensor and actuator fault channels except the rotor–effective wind speed $v_r$ and tower–top acceleration $\ddot{x}_t$, which are not considered. Sensor faults enter the measurement equation through $F_f$, and actuator faults enter the state equation through $E_f$:
\begin{align}
x(k+1) &= Ax(k) + Bu(k) + E_d d(k) + E_f f(k), \label{eq:process}\\
y(k) &= Cx(k) + Du(k) + F_d d(k) + F_f f(k). \label{eq:measurement}
\end{align}
Here, the columns of $E_f$ coincide with the affected input channels in $B$ (zeros elsewhere), and $F_f$ selects or weights the affected measured outputs in $C$.

\section{Fault Scenarios}\label{sec:fault_scenarios}

Ten datasets are used, each representing a distinct operating condition of the wind farm with mean winds from $5$ to $\SI{23}{m/s}$.  
Each dataset contains exactly nine single–fault episodes, affecting one turbine at a time. Fault durations range from $10$ to $\SI{50}{s}$; severity and onset time are varied across datasets to obtain heterogeneous yet comparable scenarios. All injections respect physical limits (e.g., $\beta\in[\beta_{\min},\beta_{\max}]$, $T_g\in[0,T_{g,\max}]$). Fault classes and magnitudes are based on typical sensor and actuator failures reported in~\cite{odgaard2013fault1,odgaard2013fault2,johnson2011development,habibi2019reliability}. Faults in wind-speed and tower-acceleration sensors are not considered in this study, as the focus is placed on pitch and torque systems, generator speed, and power measurements, which are relevant to turbine control and fault diagnosis. In particular, field-data analyses identify pitch systems as critical subsystems \cite{Walgern2023_EnergyReports_PitchReliability}. The remaining channels are selected for their role in the control and power-conversion loops, not on the basis of a universal failure-rate ranking. Nacelle anemometers sample flow disturbed by the rotor and may exhibit systematic measurement errors \cite{Astolfi2024_Energies_StaticErrors}; therefore, we do not treat wind-speed measurements as an independent fault-free diagnostic reference in this study.

The individual injections and detection outcomes for the reported $\SI{13}{m/s}$ scenario are summarized in Table~\ref{tab:fault_outcomes} in Sec.~\ref{sec:results}.

For a scalar signal $q(t)$ and its faulty counterpart $\tilde q(t)$, faults are injected on the interval $[t_0,t_1)$ according to the following generic models (outside this interval $\tilde q(t)=q(t)$):

\begin{align}
  \text{Stuck:}\quad
  \tilde q(t) &=
  \begin{cases}
  q(t), & t<t_0,\\
  \tilde q(t_0^-), & t\in[t_0,t_1),\\
  q(t), & t\ge t_1,
  \end{cases} \label{eq:fault_stuck} \\[0.5em]
  \text{Offset:}\quad
  \tilde q(t) &= q(t) - b\,\mathbf{1}_{[t_0,t_1)}(t), \label{eq:fault_offset} \\[0.5em]
  \text{Drift:}\quad
  \tilde q(t) &= q(t) + \kappa (t-t_0)\,\mathbf{1}_{[t_0,t_1)}(t), \label{eq:fault_drift} \\[0.5em]
  \text{Scaling:}\quad
  \tilde q(t) &= g\,q(t)\,\mathbf{1}_{[t_0,t_1)}(t)
  + q(t)\,\mathbf{1}_{\mathbb{R}\setminus [t_0,t_1)}(t), \label{eq:fault_gain}
\end{align}
with bias $b$, drift slope $\kappa$, and gain $g$ chosen per fault type as specified below. For a sensor stuck fault, the reported measurement is held at its pre-fault value, while the physical state may continue evolving. For an actuator stuck fault as modelled here, the input reference to the actuator dynamics is held constant, not the actuator state itself. For example, a frozen torque reference $T_c$ gives $T_g(t)=T_c+(T_g(t_0)-T_c)e^{-(t-t_0)/\tau_g}$ on the fault interval under the nominal first-order torque dynamics. The pitch angle and pitch rate likewise continue to satisfy the second-order actuator equation with a constant reference. A mechanically locked actuator would require a different fault model imposing a state constraint; it is not implied by the generic stuck-signal equation.

\paragraph{Pitch sensor faults}

Pitch sensors measure the blade angles $\beta_j(t)$ and output $\tilde{\beta}_j(t)$.  
Each fault affects a single blade $j$ and follows one of the generic models~\eqref{eq:fault_stuck}–\eqref{eq:fault_gain} with $q(t)=\beta_j(t)$. The parameter ranges are $|b|\le 3^{\circ}$, $g \equiv g_\beta \in[0.8,1.2]$, and $\kappa$ denotes the drift slope.

\paragraph{Generator sensor faults}

Generator sensors measure shaft speed $\omega_g$ and electrical power $P_{\mathrm{meas}}$, yielding $\tilde{\omega}_g$ and $\widetilde{P}_{\mathrm{meas}}$.  
Both channels share the same fault logic, using~\eqref{eq:fault_stuck}–\eqref{eq:fault_gain} with $q\in\{\omega_g,P_{\mathrm{meas}}\}$. The gains are bounded by $g_q\in[0.95,1.05]$ for $\omega_g$, $g_q\in[0.9,1.1]$ for $P_{\mathrm{meas}}$.

\paragraph{Pitch actuator faults}

The pitch actuator converts the controller reference $\beta_{\mathrm{ref}}$ into the commanded angle $\tilde{\beta}_{\mathrm{ref}}$, which is then fed to the actuator dynamics.  
Faults act on $\beta_{\mathrm{ref}}$ via the generic models~\eqref{eq:fault_stuck}–\eqref{eq:fault_gain} with $q(t)=\beta_{\mathrm{ref}}(t)$, $|b|\le 3^{\circ}$, $g \equiv g_\beta\in(0,1.2)$, and an additional dynamics fault modifies the first–order response as
\[
\dot{\tilde \beta}_{\mathrm{ref}}
  =\frac{\beta_{\mathrm{ref}}-\tilde \beta_{\mathrm{ref}}}{\tau_\beta^{\mathrm{fault}}},
  \qquad \tau_\beta^{\mathrm{fault}}>\tau_\beta.
\]

\paragraph{Generator actuator faults}

The generator torque actuator produces the effective torque $\tilde{T}_{g,\mathrm{ref}}$ from the controller reference $T_{g,\mathrm{ref}}$.  
Faults act on $T_{g,\mathrm{ref}}$ according to~\eqref{eq:fault_stuck}–\eqref{eq:fault_gain} with $q(t)=T_{g,\mathrm{ref}}(t)$ and $|b|\le \SI{1}{kNm}$, $g \equiv g_T=\text{GenActGain}$, and, in addition, a loss–of–effectiveness mode
\[
\tilde T_{g,\mathrm{ref}}(t)=e\,T_{g,\mathrm{ref}}(t), 
\qquad e\in(0,1),
\]
where $e$ denotes the remaining actuator effectiveness.

\section{Statistical change detection under uncertainty}
\label{sec:stat}

For each turbine, let $\hat x(k)$ be the observer state estimate and $\hat y(k)=C\hat x(k)+Du(k)$ the predicted output. The innovation is $e(k)=y(k)-\hat y(k)$, and the residual used for detection is $r(k)=W_0e(k)$ for the SKR-based filter or $r(k)=Ve(k)$ for the unified solution, with the respective post-filter matrices defined in Appendix~\ref{appendixA} (see \eqref{eqn:obs2} and \eqref{eqn:obs_form}). The turbine index is suppressed throughout this section.

The objective is to detect a mean shift in the residual \(r(k)\) induced
by faults, under parameter uncertainty. We consider the hypothesis test
\[
H_0:\ \mu=0\quad\text{vs.}\quad
H_1:\ \mu=
\begin{cases}
0, & i<k_0,\\
\mu_1\neq 0, & i\ge k_0,
\end{cases}
\]
with unknown change time \(k_0\) and post-change mean shift \(\mu_1\).

The random-parameter description represents epistemic uncertainty rather than physical resampling of the turbine coefficients at every time step. In particular, the coefficients of the aerodynamic $C_p$ and $C_t$ surrogates are inferred from finite information and are imperfectly known. A probability distribution over plausible coefficient vectors quantifies this uncertainty; in the implementation, it is approximated by a Gaussian distribution using the offline parameter inference described in Sec.~\ref{sec:implementation}. Even if the physical coefficients are constant during a run, different plausible coefficient vectors give different predicted trajectories and residuals. This induces a distribution of residuals and may induce temporal dependence, since the same uncertain coefficients enter successive time steps. The Gaussian approximation and the independence assumptions used for the idealized GLR law below are therefore distinct assumptions, not consequences of physical parameter variability.

A statistically valid decision rule for distinguishing $H_0$ from
$H_1$ must account for uncertainty in the model parameters, which
propagates into the residual statistics. The analysis proceeds in two steps:
(i) a first–order Taylor approximation is used to characterize the effect of
parameter perturbations on the distribution of \(r(k)\), enabling the construction
of a GLR test; (ii) second–order terms in the parameter expansion are later used to
assess departures from Gaussianity and to guide an appropriate threshold selection strategy.

\begin{lemma}[First–order delta method]
\label{lem:first_order_delta_residual}
Let \(\theta\in\mathbb R^p\) denote the vector of model parameters, with nominal
value \(\theta_0\) and perturbation \(\delta\theta:=\theta-\theta_0\). Assume
that the parameter perturbation is Gaussian,
\(\delta\theta\sim\mathcal N(0,\Sigma_\theta)\). Let \(r(k;\theta)\in\mathbb R^m\) denote the residual
at time \(k\). A first–order Taylor expansion at \(\theta_0\) yields
\begin{equation}
    r(k;\theta)\;=\;r_0(k)\;+\;J_r(k)\,\delta\theta\;+\;\mathcal O(\|\delta\theta\|^2),
    \label{eq:res_firstorder_taylor}
\end{equation}
where \(J_r(k)=\big[\partial r/\partial\theta\big]_{\theta_0}\) and, under
whitening, \(r_0(k)=0\) under \(H_0\).

If the Jacobian estimate converges
in probability to the population Jacobian, \(\widehat J_r(k)\xrightarrow{p}J_r(k)\), and
\(\delta\theta\xrightarrow{d}\mathcal N(0,\Sigma_\theta)\) (where $\xrightarrow{p}$ and $\xrightarrow{d}$ denote convergence in probability and in distribution, respectively), then by Slutsky’s
theorem \cite{van2000asymptotic},
\[
\widehat J_r(k)\,\delta\theta \;\xrightarrow{d}\; J_r(k)\,\delta\theta
\;\sim\; \mathcal N\!\big(0,\;J_r(k)\Sigma_\theta J_r(k)^\top\big).
\]
\end{lemma}
Hence, by Lemma~\ref{lem:first_order_delta_residual}, the residual is approximately zero-mean Gaussian with covariance
\begin{equation}
    \Sigma_{r}(k)=J_r(k)\,\Sigma_\theta\,J_r(k)^\top.
    \label{eq:Sigma_r_theo_first}
\end{equation}
This approximation justifies the subsequent GLR formulation and the corresponding $\chi^2$ thresholding.

\begin{definition}[Windowed GLR test for a mean shift]
\label{def:glr}
Let \(\mathcal W_k=\{k-M{+}1,\dots,k\}\) denote the most recent window.
For a candidate change time \(j\in\mathcal W_k\), let
\(N_{j,k}=k-j+1\) and freeze the covariance at \(\Sigma_r(j)\).
The generalized likelihood–ratio statistic for a change at \(j\) is\cite{basseville1993detection}
\begin{equation}
    g_j(k)
=\frac{1}{2\,N_{j,k}}
\Bigl(\textstyle\sum_{i=j}^{k} r(i)\Bigr)^{\!\top}
\Sigma_r^{-1}(j)
\Bigl(\textstyle\sum_{i=j}^{k} r(i)\Bigr).
\label{eqn:gj}
\end{equation}
Define \(J_j(k)=2\,g_j(k)\). The windowed decision statistic is
\begin{equation}
J(k)=\max_{k-M+1\le j\le k} J_j(k),
\end{equation}
and an alarm is raised when
\begin{equation}
k_a=\inf\{k: J(k)>\gamma\},
\end{equation}
where \(\gamma\) is chosen to meet a desired false–alarm rate.
The estimated change time is
\begin{equation}
\hat{k}_0=\arg\max_{\,k_a-M+1\le j\le k_a} J_j(k_a).
\label{eqn:k0}
\end{equation}
\end{definition}

\begin{proposition}[\(\chi^2\) distributions of the GLR statistic]
\label{prop:glr_chi2}
Let \(z(i)=\Sigma_r^{-1/2}(j)\,r(i)\). Under \(H_0\),
\(z(i)\stackrel{i.i.d.}{\sim}\mathcal N(0,I)\), and
\[
J_j(k)=\frac{1}{N_{j,k}}\left\|\sum_{i=j}^{k} z(i)\right\|^2
\sim \chi^2(m).
\]
Under \(H_1\) with mean shift \(\mu_1\),
\(\sum_{i=j}^{k}z(i)\sim\mathcal N(N_{j,k}\mu_z, N_{j,k}I)\) with 
\(\mu_z=\Sigma_r^{-1/2}(j)\mu_1\), giving
\[
J_j(k)\sim\chi^2\bigl(m,\lambda\bigr),\qquad
\lambda=N_{j,k}\mu_1^\top\Sigma_r^{-1}(j)\mu_1.
\]
Thus, a threshold \(\gamma\) corresponding to false-alarm probability \(\alpha\)
is obtained from the \((1-\alpha)\)-quantile of the central \(\chi^2(m)\) law.
\end{proposition}

The residual Jacobian \(J_r(k)\) depends on the
sensitivities of the plant and filter blocks to the uncertain parameters.
To compute these sensitivities in real time, first–order perturbations are
propagated through the observer dynamics on a per-parameter basis.

Let \(J_Z^{\theta_j}:=[\partial Z/\partial \theta_j]_{\theta_0}\) denote the
Jacobian of any plant or filter matrix \(Z\in\{A,B,C,D,P,L,W\}\) with
respect to parameter \(\theta_j\). Closed-form expressions for the
first-order sensitivities \(J_P^{\theta_j}\), \(J_L^{\theta_j}\), and \(J_W^{\theta_j}\)
with respect to the plant parameters are obtained from the first-order perturbation maps \eqref{eqn:P1}, \eqref{eqn:L1} and \eqref{eqn:W1} derived in Appendix~\ref{appendixB} for the SKR-based filter described in Appendix~\ref{appendixA}. The resulting first-order filter sensitivities are then propagated to the state, output, and residual
signals to obtain $J_{\hat x(k+1)}^{\theta_j}$, $J_{\hat y(k)}^{\theta_j}$ and $J_{r(k)}^{\theta_j}$, respectively, via the first–order recursions as in \eqref{eq:dx1}, \eqref{eq:dy1} and \eqref{eq:dr1}.

Stacking the per-parameter residual Jacobians gives
\[
J_r(k)=
\begin{bmatrix}
J_{r(k)}^{\theta_1} &
\cdots &
J_{r(k)}^{\theta_p}
\end{bmatrix},
\]
which fully determines the first–order covariance
\(\Sigma_r(k) = J_r(k)\Sigma_\theta J_r(k)^\top\) in
\eqref{eq:Sigma_r_theo_first}.

\begin{definition}[Fisher information over a window]
\label{def:fisher_residual}
Let \(R\) denote the (noise–only) residual covariance under \(H_0\) at
\(\theta_0\), and let \(\mathcal W\) be a time window. The Fisher information matrix
associated with the residual sensitivities \(\{J_{r(k)}^{\theta_j}\}\) over \(\mathcal W\) is
\begin{equation}
\big[\mathcal I(\theta_0;\mathcal W)\big]_{j\ell}
=\sum_{k\in\mathcal W} \big(J_{r(k)}^{\theta_j}\big)^\top\,R^{-1}\,J_{r(k)}^{\theta_\ell},
\end{equation}
for \(j,\ell=1,\dots,p\). The matrix \(\mathcal I(\theta_0;\mathcal W)\) is symmetric
positive semidefinite.
\end{definition}

Diagonal entries of \(\mathcal I(\theta_0;\mathcal W)\) rank the per–parameter influence
on the residual, while off–diagonal entries reflect sensitivity coupling and
potential lack of practical identifiability.

We now examine the effect of retaining second–order terms in the Taylor
expansion of the residual and the implications for threshold calibration.

\begin{lemma}[Mixed first/second–order delta method]
\label{lem:second_order_residual}
Let \(\theta\in\mathbb R^p\) denote the vector of model parameters, with nominal
value \(\theta_0\) and perturbation \(\delta\theta:=\theta-\theta_0\). Assume
\(\delta\theta\sim\mathcal N(0,\Sigma_\theta)\). Let \(r(k;\theta)\in\mathbb{R}^m\)
denote the residual at time \(k\). A second-order Taylor
expansion of \(r(k;\theta)\) about \(\theta_0\) yields
\begin{equation}
    r(k;\theta)
    {\approx}
    r_0(k)
    {+} J_r(k)\,\delta\theta
    {+} \tfrac12\,
      \mathcal{H}_r(k)[\delta\theta,\delta\theta]
    {+} \mathcal O(\|\delta\theta\|^3),
    \label{eq:res_secondorder_vector}
\end{equation}
where \(J_r(k)=\left[\partial r / \partial\theta\right]_{\theta_0}\)
is the residual Jacobian and
\(\mathcal{H}_r(k)[\cdot,\cdot]\) denotes
\begin{equation}
\mathcal{H}_r(k)[\delta\theta,\delta\theta]
=
\begin{bmatrix}
\delta\theta^\top H_{r,1}(k)\,\delta\theta \\
\vdots\\
\delta\theta^\top H_{r,m}(k)\,\delta\theta
\end{bmatrix},
\end{equation}
with \(H_{r,\ell}(k)\in\mathbb{R}^{p\times p}\) the Hessian of the
\(\ell\)-th component of \(r(k;\theta)\) evaluated at \(\theta_0\).

Under these assumptions, \(r(k;\theta)\) is approximately the sum of a
Gaussian term and a vector of quadratic forms in Gaussian variables. Each scalar residual component is a linear-plus-quadratic function of a Gaussian vector and can be represented by a generalized $\chi^2$ law (including a possible independent Gaussian term). A quadratic statistic of the full second-order residual generally contains terms of degree up to four in $\delta\theta$ and need not follow such a law.
\end{lemma}

Under Lemma~\ref{lem:second_order_residual}, tail probabilities and quantiles
for such generalized \(\chi^2\) laws can, in principle, be computed numerically,
e.g., via the Imhof or Davies algorithms
\cite{imhof1961computing,davies1980distribution}.

Second–order terms enter the uncertainty propagation through the plant
matrices and the filter maps constructed from them. Let the plant blocks be
collected in \(\mathcal D(\theta):=\{A(\theta),B(\theta),C(\theta),D(\theta)\}\),
and the filter blocks in \(\mathcal F(\theta):=\{P(\theta),L(\theta),W(\theta)\}
=\Phi\big(\mathcal D(\theta)\big)\), where \(\Phi\) denotes the deterministic map
from plant data to the Riccati solution, observer gain, and post-filter.

Using the series convention from Appendix~\ref{appendixB}, every variable
\(Z \in \mathcal D(\varepsilon)\cup\mathcal F(\varepsilon)\) can be expressed
using the book-keeping scalar \(\varepsilon\) as
\begin{equation}
    Z(\varepsilon)
    = Z_0 + \varepsilon Z_1 + \varepsilon^2 Z_2 + \mathcal O(\varepsilon^3),
\end{equation}
where \(Z_1\) and \(Z_2\) denote the first- and second-order perturbations with
respect to \(\delta\theta\).

At the plant level, the coefficients are given by the usual Taylor maps with
respect to \(\theta\) at \(\theta_0\),
\begin{align}
X_1 &= \Big[\tfrac{\partial X}{\partial\theta}\Big]_{\theta_0}\,\delta\theta,
\label{eq:X1_plant} \\[2pt]
X_2 &= \tfrac12\,\delta\theta^\top
      \Big[\tfrac{\partial^2 X}{\partial\theta^2}\Big]_{\theta_0}
      \delta\theta,
\label{eq:X2_plant}
\end{align}
for each \(X\in\mathcal D(\theta)\). At the filter level, the coefficients
\(Y_1,Y_2\) for each \(Y\in\mathcal F(\theta)\) follow from the chain rule
applied to the map \(\Phi\),
\begin{align}
Y_1 &= \mathcal L_\Phi[\mathcal D_0](\{X_1\}),
\label{eq:Y1_filter}\\[2pt]
Y_2 &= \mathcal L_\Phi[\mathcal D_0](\{X_2\})
      +\tfrac12\,\mathcal L_\Phi^{(2)}[\mathcal D_0](\{X_1,X_1\}),
\label{eq:Y2_filter}
\end{align}
The first equation propagates the linear plant perturbations through the local derivative of the filter-design map. The second separates two sources of curvature: the response to the intrinsic second-order plant perturbations $X_2$, and the curvature of the filter-design map itself acting on pairs of first-order perturbations $X_1$. Thus, even when $X_2$ is neglected, the second term can be nonzero. Here \(\mathcal L_\Phi\) and \(\mathcal L_\Phi^{(2)}\) denote the first and
second Fréchet derivatives of \(\Phi\). In particular, Appendix~\ref{appendixB}
provides the explicit expressions for the SKR-based filter, yielding \(P_1,P_2\) in
\eqref{eqn:Delta_P}--\eqref{eqn:P2}, \(L_1,L_2\) in
\eqref{eqn:L0}--\eqref{eqn:L2}, \(W_1,W_2\) in
\eqref{eqn:W0}--\eqref{eqn:W2}, and the corresponding
residual perturbations \(r_1(k),r_2(k)\) in
\eqref{eq:dr1}--\eqref{eq:dr2}. The same derivation applies to the
unified solution, also described in Appendix~\ref{appendixA}, by replacing \((B,D)\!\to\!(E_d,F_d)\) and \(W\!\to\!V\), and by
interpreting \(Q_r=E_dE_d^\top\), \(R_r=F_dF_d^\top\), \(S_r=E_dF_d^\top\) in
the standard form of the Riccati equation in~\eqref{eq:DARE-standard}.

\begin{remark}[On generalized \(\chi^2\) calibration]
Computing quantiles of generalized \(\chi^2\) distributions is
computationally expensive and becomes infeasible for large-scale systems.
A distribution–free calibration is therefore preferred; it also has the
advantage of remaining robust to departures from the quadratic approximation
due to unmodelled dynamics.
\end{remark}

A distribution-free calibration is constructed to target, for a specified
per-time false-alarm rate $\alpha_{\mathrm{total}}\in(0,1)$, that $\mathbb{P}\!\left\{ J(k)\ge h \right\} \le \alpha_{\mathrm{total}}$ under healthy operation. At a fixed time \(k\), the decision involves the \(M\) statistics
\(\{J_j(k)\}_{j=k-M+1}^k\), defined on overlapping windows.

\begin{definition}[Family-wise error control for multiple tests]
\label{def:fwer}
Consider \(M\) hypothesis tests with test statistics
\(T_1,\dots,T_M\), each compared to a threshold \(\gamma\).
A null hypothesis is the hypothesis of no change in the residual mean ($H_0:\mu=0$ in this application). A true null is one for which no such change has occurred; rejecting it constitutes a false alarm. Under healthy operation all candidate-window null hypotheses are true. The family-wise error rate (FWER) is
\[
\mathrm{FWER} = \mathbb{P}\{\text{at least one true null is rejected}\}.
\]
A multiple-comparison correction maps a desired global level
\(\alpha_{\mathrm{total}}\) to a per-test level \(\alpha_{\mathrm{single}}\)
such that
\[
\mathrm{FWER}\le \alpha_{\mathrm{total}}.
\]
\end{definition}

Two classical corrections are the Bonferroni and Sidák rules
\cite{hsu1996multiple}:

\begin{definition}[Bonferroni and Sidák corrections]
\label{def:bonf_sidak}
Let \(\alpha_{\mathrm{total}}\in(0,1)\).
\begin{itemize}
\item \textbf{Bonferroni:}  
      \(\alpha_{\mathrm{single}} = \alpha_{\mathrm{total}}/M\).  
      This follows from the union bound and requires no independence.
\item \textbf{Sidák:}  
      \(\alpha_{\mathrm{single}} = 1-(1-\alpha_{\mathrm{total}})^{1/M}\).  
      This is exact under independence.
\end{itemize}
\end{definition}

Bonferroni can be conservative for strongly dependent overlapping windows. Sidák with an estimated effective multiplicity is a heuristic under dependence, not an exact FWER guarantee. Likewise, substituting sample moments for population moments in the inequalities below does not establish a finite-sample guarantee.

\begin{definition}[Effective multiplicity]
\label{def:Meff}
Dependence among the \(M\) candidates is accounted for by replacing \(M\)
with an effective multiplicity \(M_{\mathrm{eff}}\), estimated from a
healthy calibration segment using the Li--Ji rule \cite{li2005adjusting}.
Let the values \(\{J_j(k)\}\) be computed for many calibration times
and stacked into a matrix whose columns index the window starts \(j\).
Let \(\{\lambda_m\}_{m=1}^M\) denote the eigenvalues of the sample
correlation matrix across columns. The effective multiplicity is
\[
M_{\mathrm{eff}} = \sum_{m=1}^{M} \min(\lambda_m,1).
\]
\end{definition}

With \(M_{\mathrm{eff}}\) in place of \(M\), the Sidák correction yields
the per-test level
\begin{equation}
  \alpha_{\mathrm{single}}
  = 1 - (1-\alpha_{\mathrm{total}})^{1/M_{\mathrm{eff}}}.
  \label{eq:alpha_single}
\end{equation}

Empirical mean and variance of the healthy statistics \(\{J_j(k)\}\) are
estimated from the same calibration segment:
\begin{equation}
  \hat\mu_J = \frac{1}{L}\sum_{\ell=1}^{L} J_\ell,
  \qquad
  \hat\sigma_J^2 = \frac{1}{L-1}\sum_{\ell=1}^{L}
  (J_\ell-\hat\mu_J)^2,
  \label{eq:mu_sigma_hat}
\end{equation}
where \(L\) is the number of pooled samples.

\begin{definition}[Distribution-free per-test thresholds]
\label{def:df_thresholds}
For a per-test level \(\alpha_{\mathrm{single}}\), two one-sided
distribution-free thresholds for \(J_j(k)\) are obtained from:

\begin{itemize}
  \item \textbf{Markov's inequality}  
    For any nonnegative \(X\),
    \(\mathbb{P}\{X\ge\gamma\} \le \mathbb{E}[X]/\gamma\).  
    Applied to \(X=J_j(k)\),
    \[
    \gamma_{\mathrm{Markov}}
      = \frac{\hat\mu_J}{\alpha_{\mathrm{single}}},
      \qquad
      h_{\mathrm{Markov}}=\frac{\gamma_{\mathrm{Markov}}}{2}.
    \]

  \item \textbf{Cantelli's inequality (one-sided Chebyshev)}  
    For any real-valued \(X\) with mean \(\mu\) and variance \(\sigma^2\),
    \(\mathbb{P}\{X-\mu\ge t\}\le \sigma^2/(\sigma^2+t^2)\).  
    Solving for \(t\) with \(\alpha_{\mathrm{single}}=\sigma^2/(\sigma^2+t^2)\)
    gives
    \[
    \gamma_{\mathrm{Cantelli}}
      = \hat\mu_J
        + \hat\sigma_J \sqrt{\frac{1-\alpha_{\mathrm{single}}}{\alpha_{\mathrm{single}}}},
      \hspace{2mm}
      h_{\mathrm{Cantelli}}=\frac{\gamma_{\mathrm{Cantelli}}}{2}.
    \]
\end{itemize}
\end{definition}

\section{Implementation}\label{sec:implementation}

We use benchmark time series at a data rate \(T_s^{\mathrm{data}}=0.1\,\mathrm{s}\) and a control/filtering rate \(T_s=0.0125\,\mathrm{s}\) over a horizon of \(900\,\mathrm{s}\).
All measured channels (rotor/generator speed, generator torque, blade pitches, nacelle acceleration, wind speed at the rotor, references, and farm power) are interpolated from \(T_s^{\mathrm{data}}\) to \(T_s\) via shape-preserving cubic Hermite interpolation. Seven turbines are processed in parallel; measurement and input matrices are assembled by stacking per–turbine rows and later addressed by fixed index patterns. A steady operating point at \(v_r=13\,\mathrm{m/s}\) is obtained by solving the closed-loop equilibrium equations. All matrices and signals are normalized about this operating point with diagonal scalings \(N_x, N_u, N_y\) (state, input, output), yielding dimensionless state--space data
\((A,B,C,D)\).
Initial conditions, operating points \((x_\mathrm{op},u_\mathrm{op},y_\mathrm{op})\), and the full data streams are scaled consistently. The continuous-time Jacobians are discretized by forward Euler.

The system matrices affected by the considered uncertainty sources are $A$ and $C$. We construct two sources of model uncertainty:

\paragraph{Operating–wind variability.}
Let the steady relations be given by \(F(z,v_r)=0\) with
\(z=[\omega_r,\ T_g,\ \beta,\ \omega_g^{\mathrm{err}},\ x_t,\ \dot x_t]^\top\).
Form the Jacobians \(J_z=\tfrac{\partial F}{\partial z}\) and \(\tfrac{\partial F}{\partial v_r}\).
By the implicit–function theorem,
\begin{equation}
\begin{split}
    \frac{\mathrm d z}{\mathrm d v_r}\Big|_{\mathrm{op}}
    &= -\,J_z^{-1}\,\frac{\partial F}{\partial v_r}\Big|_{\mathrm{op}}\\
    &=\ \Big[\tfrac{\mathrm d\omega_r}{\mathrm d v_r},\ \tfrac{\mathrm d T_g}{\mathrm d v_r},\ \tfrac{\mathrm d \beta}{\mathrm d v_r},\ \tfrac{\mathrm d \omega_g^{\mathrm{err}}}{\mathrm d v_r},\ \tfrac{\mathrm d x_t}{\mathrm d v_r},\ \tfrac{\mathrm d \dot x_t}{\mathrm d v_r}\Big]^\top
\end{split}
\end{equation}
Injecting these total derivatives into the chain rule gives the sensitivities of \(A\) and \(C\) with respect to wind
\begin{align*}
\frac{\mathrm d A}{\mathrm d v_r}
&= \frac{\partial A}{\partial v_r}
+ \sum_{q\in\{\omega_r,\beta,\omega_g^{\mathrm{err}},x_t,\dot x_t\}}
\frac{\partial A}{\partial q}\,\frac{\mathrm d q}{\mathrm d v_r},\\
\frac{\mathrm d C}{\mathrm d v_r}
&= \frac{\partial C}{\partial v_r}
+ \sum_{q\in\{\omega_r,\beta,\omega_g^{\mathrm{err}},x_t,\dot x_t\}}
\frac{\partial C}{\partial q}\,\frac{\mathrm d q}{\mathrm d v_r}.
\end{align*}
With \(v_r\sim\mathcal N(v_{r,\mathrm{op}},\sigma_{v_r}^2)\) and \(\sigma_{v_r}=0.5\,\mathrm{m/s}\), we form the perturbations due to slight variations in wind operating point
\[
\Delta A^{(\mathrm{wind})}=\frac{\mathrm d A}{\mathrm d v_r}\,\sigma_{v_r},
\qquad
\Delta C^{(\mathrm{wind})}=\frac{\mathrm d C}{\mathrm d v_r}\,\sigma_{v_r}.
\]

\paragraph{$C_p/C_t$ parametric model.}
Let $\theta=\big[c_1,\ldots,c_5,\ K_{00},\ldots,K_{22}\big]^\top$, $\theta\sim\mathcal N(\theta_0,\Sigma_\theta)$, where \(\Sigma_\theta\) is estimated offline via NUTS Hamiltonian Monte-Carlo (HMC) in \textsc{Turing.jl} \cite{gomez2025reference}.
Assuming that the parameters are a priori uncorrelated, retain only the variances
\(\sigma_i^2 = [\Sigma_\theta]_{ii}\) and form the parameter–driven perturbations
\[
\Delta A^{(\mathrm{par})}=\sum_{i=1}^p \frac{\partial A}{\partial\theta_i}\,\sigma_i,
\qquad
\Delta C^{(\mathrm{par})}=\sum_{i=1}^p \frac{\partial C}{\partial\theta_i}\,\sigma_i.
\]

Then, the plant–layer perturbations used in the uncertainty propagation are
\begin{align}
\Delta A &= \Delta A^{(\mathrm{par})}+\Delta A^{(\mathrm{wind})},\\
\Delta C &= \Delta C^{(\mathrm{par})}+\Delta C^{(\mathrm{wind})}.
\end{align}
where $\Delta A$ and $\Delta C$ correspond to $A_1$ and $C_1$ in the notation used in Appendix~\ref{appendixB}.

For both the SKR orthogonal–projection filter and the unified solution, $(A_1,C_1)$ are propagated through the Riccati solutions, gains, and post–filters up to second order following the procedure shown in Appendix~\ref{appendixB}. Intrinsic second-order terms of the uncertain system matrices ($A_2$, $C_2$) are neglected for simplicity. Only the second-order contributions arising from products of first-order terms are retained.
The directional approximation of the time–varying residual covariance used in the subsequent statistical change detection stage is
\begin{equation}
\Sigma_r(k)=r_1(k)r_1(k)^\top + R + R_{\mathrm{np}},
\end{equation}
where $r_1(k)$ is the first-order residual perturbation as in \eqref{eq:dr1}, $R$ is the sensor noise covariance and $R_{\mathrm{np}}$ is a heuristic covariance added to account for non-parametric (unmodelled) uncertainties, as detailed below. 
The propagated aerodynamic uncertainty is negligible for the actuator-related residuals $(r_{\beta_1},r_{\beta_2},r_{\beta_3},r_{T_g})$. However, healthy data contain occasional spikes that may result from unmodelled actuator dynamics, such as hysteresis or rate limits. Using only the propagated uncertainty and sensor noise can therefore give these residuals excessive weight in the GLR. The diagonal augmentation $R_{\mathrm{np}}$ provides a variance floor for the affected channels.

For the actuator channels, the added variance is half the nominal residual variance estimated from healthy data, i.e., $[R_{\mathrm{np}}]_{\ell\ell}\approx 0.5\,\hat v_\ell$, where $\hat v_\ell$ denotes the healthy residual variance of channel $\ell$. This corresponds to an added standard deviation of about $0.71\sqrt{\hat v_\ell}$. The factor $0.5$ is a heuristic choice. A reproducible calibration procedure is to estimate $\hat v_\ell$ from a specified fault-free segment using the sample variance about the sample mean, apply the stated factor, and keep the augmentation fixed during fault evaluation. The variance reflects total observed variability, so this rule should not be interpreted as an independent estimate of the unmodelled contribution alone.

The outer product $r_1(k)r_1(k)^\top$ has rank at most one and is therefore singular for a multidimensional residual. The inclusion of the sensor noise covariance $R$ ensures that $\Sigma_r(k)$ is positive definite if $R$ is positive definite. The diagonal augmentation further increases the variance in selected channels, reducing the risk of excessive weighting in the GLR. Although positive definiteness ensures invertibility, it does not guarantee good numerical conditioning. For numerical evaluation of the GLR, solving a linear system with $\Sigma_r(j)$, for example using a Cholesky factorization when positive definite, is preferable to explicitly forming its inverse.


Furthermore, per-parameter sensitivities \((J_A^{\theta_j},J_C^{\theta_j})\) are also computed and propagated through the filters to obtain \(J_{r(k)}^{\theta_j}\), which is used to compute the Fisher information matrix and perform a parameter sensitivity analysis. Regarding the second-order terms, the residual Jacobians and Hessians are used to reformulate the residual expansion into a quadratic form in Gaussian variables, and the generalized $\chi^2$ toolbox\cite{das2025new} is employed to convert it to an equivalent generalized $\chi^2$ distribution and compute the 99$\%$ lower and upper quantiles.
These second-order bounds are used only to compare uncertainty envelopes. The GLR test uses the first-order covariance approximation defined above, together with the sensor noise and non-parametric covariance terms.

Lastly, for each time~$k$, the statistic $g_j(k)$ in~\eqref{eqn:gj} is evaluated for all retrospective candidates $j\in{k-M+1,\dots,k}$ with $M=50$. Whenever the decision variable $J(k)$ crosses the threshold, the change-time estimate $\hat k_0$ is obtained as in~\eqref{eqn:k0} by selecting the maximizing $j$ in the same window. Thresholds are computed from the distribution-free calibration in the end of Sec.~\ref{sec:stat}: a healthy segment of $600\,s$ is used to pool the windowed statistics ${J_j(k)}$ across $(j,k)$ and to compute $\hat\mu_J$ and $\hat\sigma_J$; the dependence induced by overlapping windows is summarized by an effective number of tests $M_{\mathrm{eff}}$ via the Li--Ji eigenvalue rule; a per-time false-alarm target $\alpha_{\mathrm{total}}=10^{-3}$ is then mapped to a per-candidate level using Sidák with $M_{\mathrm{eff}}$ in \eqref{eq:alpha_single}; and the Cantelli threshold is finally set by Def.~\ref{def:df_thresholds}. The implementation produces one threshold per turbine and, for deployment, a single farm-wide threshold given by the maximum Cantelli value across turbines.

\section{Results and Discussion}\label{sec:results}

\subsection{Uncertainty bounds}\label{subsec:bounds}
\begin{figure}
    \centering
    \includegraphics[width=0.48\textwidth]{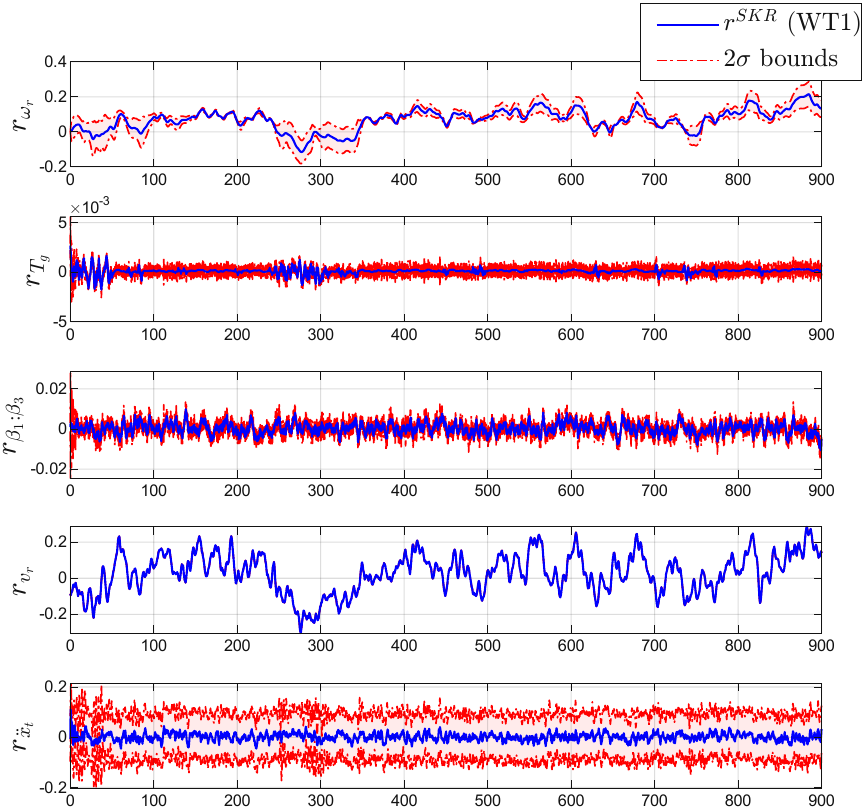}
    \caption{SKR residuals for the 1\textsuperscript{st} wind turbine and $\SI{13}{\meter/\second}$ mean wind speed under no faults. The residual covariance $\Sigma_r(k)$ is used to define 1st order uncertainty bands.}
    \label{fig:residuals_WT1_SKR_1st}
\end{figure}

\begin{figure}
    \centering
    \includegraphics[width=0.48\textwidth]{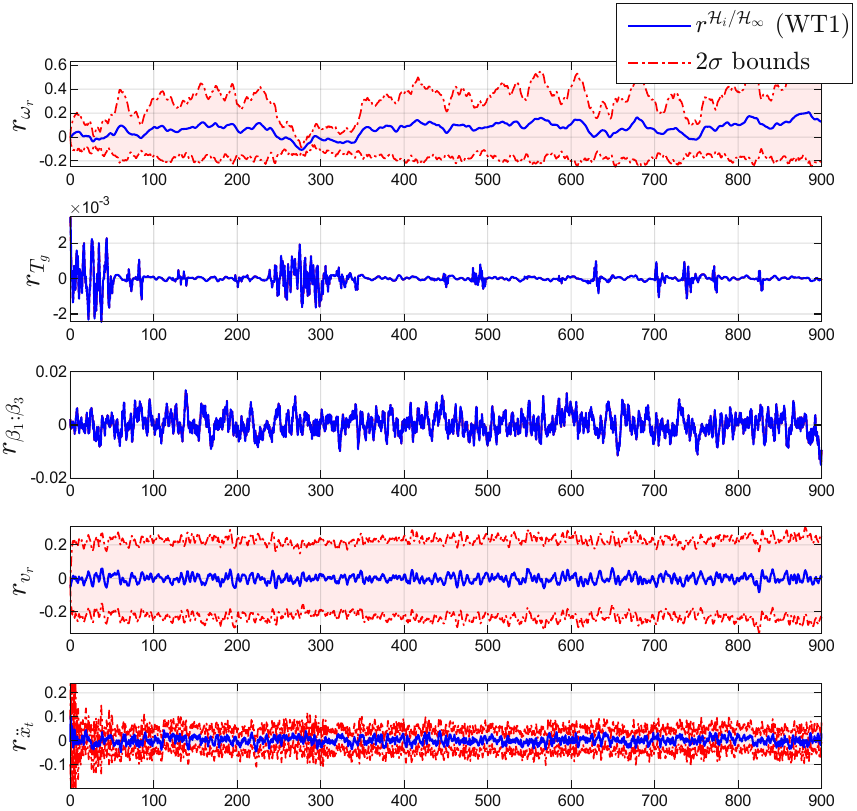}
    \caption{Unified solution residuals for the 1\textsuperscript{st} wind turbine and $\SI{13}{\meter/\second}$ mean wind speed under no faults. The residual covariance $\Sigma_r(k)$ is used to define 1st order uncertainty bands.}
    \label{fig:residuals_WT1_Hi_Hinf_1st}
\end{figure}

The plots shown in Figs.~\ref{fig:residuals_WT1_SKR_1st}--\ref{fig:residuals_WT1_Hi_Hinf_1st} shows residual time series together with $2\sigma$ uncertainty envelopes computed from the propagated residual covariance $\Sigma_r(k)$, for both the SKR orthogonal–projection filter and the unified solution. Note that the heuristic covariance contributions associated with actuator-related residuals are not included in these figures; the objective is to show only the variance bands related to parametric uncertainty.

Consider the DARE in standard form \eqref{eq:DARE-standard} with \(A_r=A^\top\) and \(B_r=C^\top\).
The structural difference between the SKR and the unified solution lies in $Q_r$; $Q_r = BB^\top$ for the SKR and $Q_r=E_dE_d^\top$ for the unified solution, as $R_r=I + DD^\top=I$, $S_r = BD^\top = 0$ for the SKR, and $R_r=F_dF_d=I$, $S_r = E_dF_d^\top = 0$ for the unified solution. Since \(E_d\) is nonzero only in the wind states \(v_m,v_t\) (rows 9–10), \(Q_r\) excites wind directions in the unified solution; \(B_d\) is nonzero only in the torque and pitch actuator channels (row 2 and rows 6–8), so \(Q_r\) only excites actuator directions in the SKR.

In the unified solution, the wind exogenous input is taken into account through $E_d,F_d$, so the Riccati equation allocates variance in the wind subspace and the gain $L$ uses the wind measurement to form a whitened innovation. As a result, the sixth residual $r_{v_r}$ is the whitened error associated with the wind channel. When uncertainty is propagated through the Riccati equation, the resulting perturbations of $P,L,V$ directly affect this residual, because the Riccati solution actively couples the uncertain dynamics to the wind–innovation channel. This is why $r_{v_r}$ shows a considerable uncertainty band in the unified solution case, which also increases the uncertainty bands in $r_{\omega_r}$ due to aerodynamic coupling.

The SKR-based filter behaves differently. Since the wind disturbance is not part of the noise model used in its Riccati equation, the filter does not attempt to explain or whiten the wind measurement. The corresponding gain entries are therefore zero in the wind direction, and the residual is the wind signal itself rather than a whitened innovation. When the same uncertainty is propagated through the SKR Riccati equation, the induced variations in $P,L,W$ do not reach this residual, because the SKR structure does not couple the uncertain dynamics to the wind channel. Hence the parametric contribution to the uncertainty of $r_{v_r}$ is effectively zero in the SKR case.

In summary, the unified solution produces a whitened wind residual with significant parametric uncertainty, while the SKR-based filter yields an unwhitened wind residual that reflects the actual wind and carries no parametric uncertainty.
For the fault scenarios considered here, which do not involve the wind or tower--acceleration sensors, the wind--speed residual $r_{v_r}$ (and similarly $r_{\ddot{x}_r}$) can be excluded from the test statistic; otherwise, the SKR-based filter would not be adequate in this setting.

\begin{figure}
    \centering
    \includegraphics[width=0.48\textwidth]{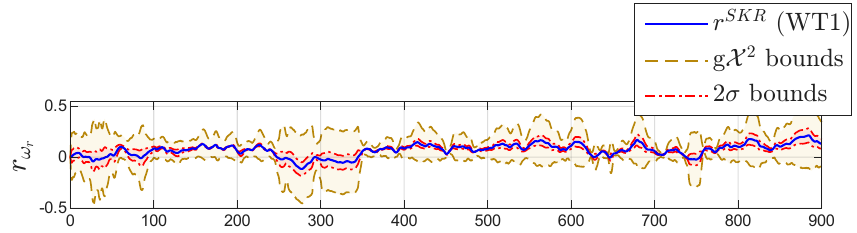}
    \caption{SKR rotor speed residual $r_{\omega_r}$ for the 1\textsuperscript{st} wind turbine and $\SI{13}{\meter/\second}$ mean wind speed under no faults. Both 1st and 2nd order (Gaussian and generalized $\chi^2$) bands are plotted.}
    \label{fig:residuals_WT1_SKR_2nd}
\end{figure}

\begin{figure}
    \centering
    \includegraphics[width=0.48\textwidth]{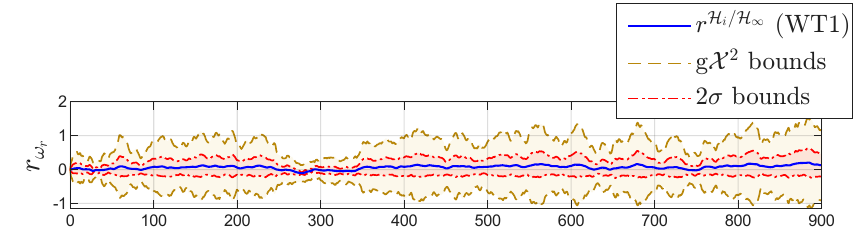}
    \caption{Unified solution rotor speed residual $r_{\omega_r}$ for the 1\textsuperscript{st} wind turbine and $\SI{13}{\meter/\second}$ mean wind speed under no faults. Both 1st and 2nd order (Gaussian and generalized $\chi^2$) bands are plotted}
    \label{fig:residuals_WT1_Hi_Hinf_2nd}
\end{figure}

The plots shown in Figs.~\ref{fig:residuals_WT1_SKR_2nd}--\ref{fig:residuals_WT1_Hi_Hinf_2nd} show a comparison between the rotor speed residual bounds computed under Gaussian assumption, and those computed considering second-order terms via the $99\%$ quantiles of a generalized $\chi^2$ distribution.

Note that, in the SKR filter, $L$ and $W$ introduce a slight cross-channel mixing in the innovation space. Consequently, the residuals associated with actuator outputs are slightly influenced by aerodynamic-related residuals, and with it the uncertainty bands associated to them. That is why a small uncertainty band still appears in $r_{T_g}$ and $r_{\beta}$ in the SKR case, even though those residuals are not affected by uncertainty in the aerodynamics. In contrast, the unified solution preserves the structural independence of actuator channels and maintains well-decoupled residuals.

\subsection{Sensitivity analysis}

This subsection reports two complementary summaries of how the residuals respond to uncertainty in $C_p/C_t$ model parameters, focusing on quantities that remain informative despite the strong dynamic–range imbalance observed across parameters.

\begin{figure}
  \centering
  \subfloat{%
    \includegraphics[width=.46\linewidth]{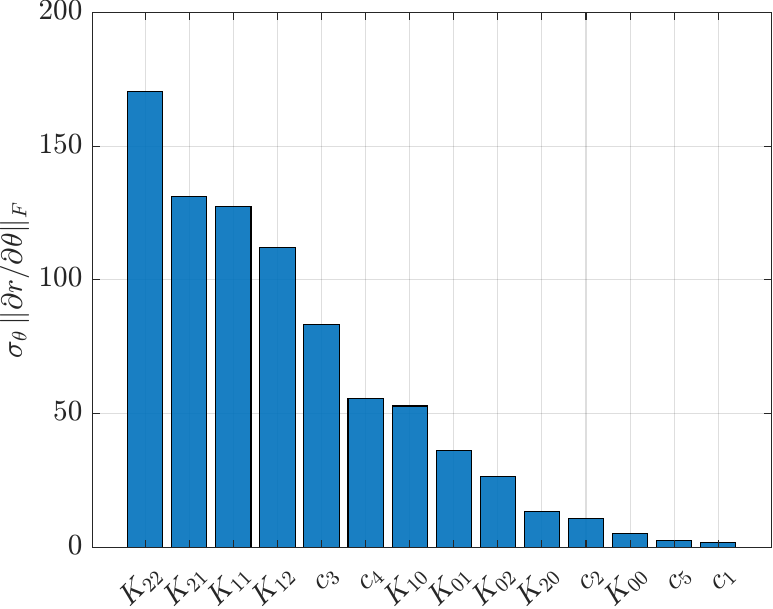}%
  }\hfill
  \subfloat{%
    \includegraphics[width=.47\linewidth]{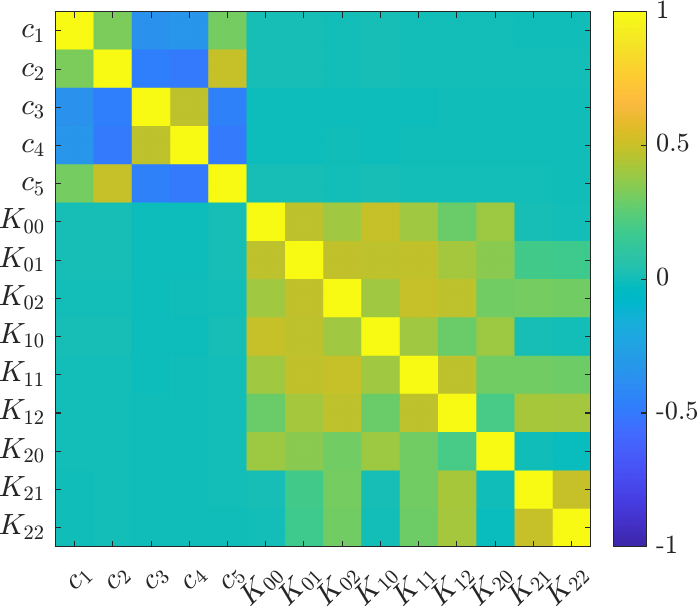}%
  }
  \caption{Standarized residual sensitivity per $1\sigma$ parameter change (left); FIM correlation matrix after diagonal normalization (right).}\label{fig:FIM}
\end{figure}

\paragraph{Standardized sensitivity}
For each parameter \(\theta_j\), the first–order residual tangent \(s_j(k)=\partial r(k)/\partial\theta_j\) is accumulated into $S_j=\bigl[s_j(1)\ \cdots\ s_j(N)\bigr]$. To account for unequal prior/posterior variances, the total effect is scaled by the parameter standard deviation, producing the score
\begin{equation}
\text{score}_j=\sigma_{\theta_j}\,\|S_j\|_F.
\end{equation}
This score quantifies the aggregate change in the residuals that would result from a \(1\sigma\) perturbation in \(\theta_j\). The resulting bar plot shown in Fig.~\ref{fig:FIM} (left) is comparable across parameters and highlights practically influential directions even when the unscaled norms are dominated by a single, extremely sensitive parameter.

\paragraph{Fisher–information correlation map}
Let \(S(k)\) stack the tangents columnwise at time \(k\) and define the time–aggregated Fisher information $F=\sum_{k=1}^{N} S(k)^\top R^{-1} S(k)$. To remove scale effects and emphasize structural dependencies, normalize:
\begin{equation}
    \mathrm{CorrFIM}=D^{-1/2}\,F\,D^{-1/2}, \qquad D=\mathrm{diag}(F).
\end{equation}
The heatmap of \(\mathrm{CorrFIM}\) shown in Fig.~\ref{fig:FIM} (right) highlights coupling between parameters independently of their individual information content: values near \(+1\) indicate strong positive co-influence on the residuals, values near \(-1\) indicate oppositely directed effects, and values near \(0\) indicate near-orthogonal (well-separated) effects.

\subsection{Decision functions under injected faults}

\begin{figure}
\centering
\begin{tikzpicture}

  \node[anchor=south west] (base) at (0,0)
    {\includegraphics[width=0.7\columnwidth]{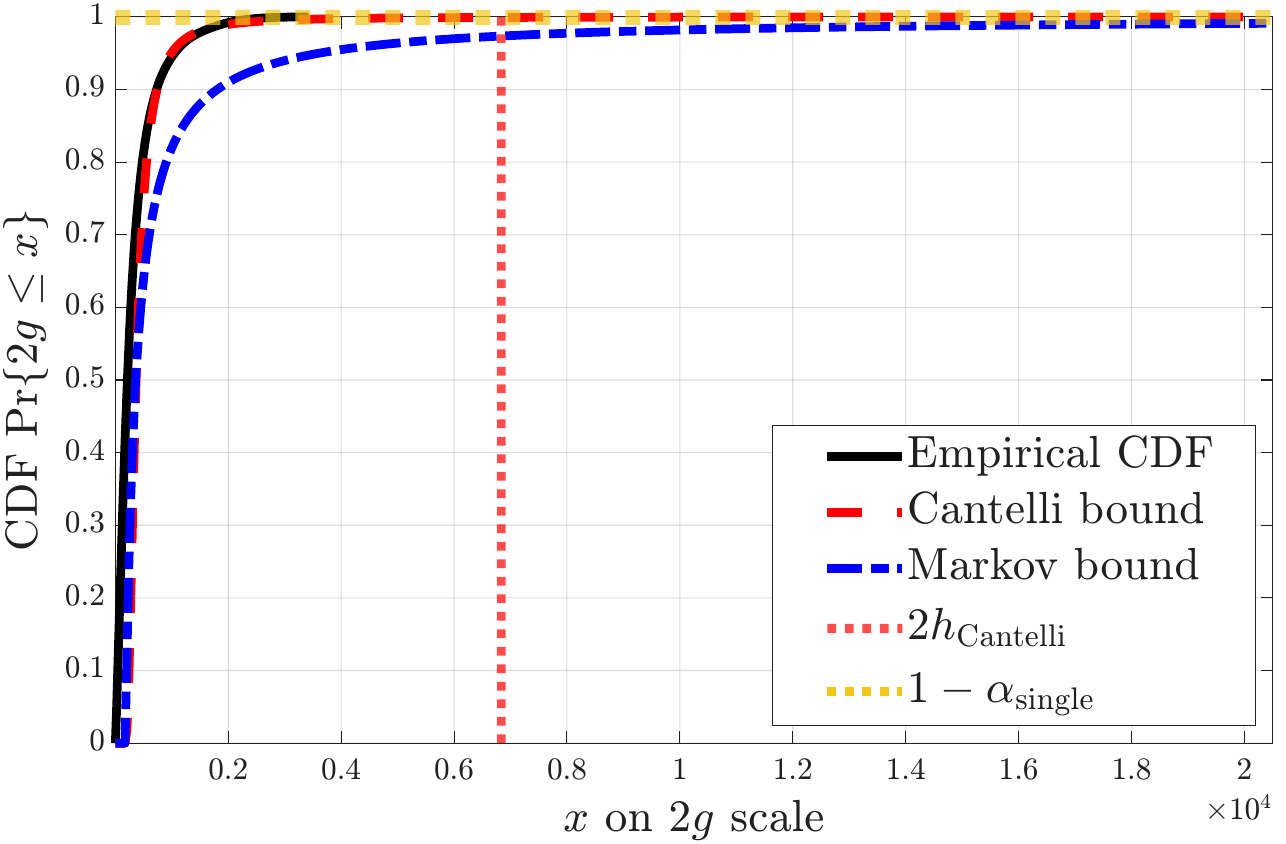}};

  \node[anchor=north east, inner sep=0pt] (zoom) at
    ([xshift=-3mm,yshift=-4mm] base.north east)
    {\includegraphics[width=0.35\columnwidth]{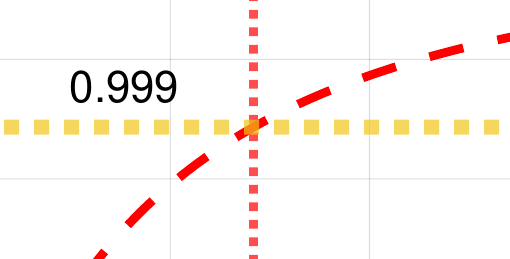}};

  \draw[dashed, black, line width=0.6pt]
    (zoom.south west) rectangle (zoom.north east);

\end{tikzpicture}
\caption{Empirical CDF of $J_j(k)=2g_j(k)$ (black) with Cantelli (red, dashed) and Markov (blue, dash–dot) bounds. Vertical red line: $2\cdot\hspace{0.5mm}$Cantelli threshold. Horizontal yellow line: $1-\alpha_{\text{single}}$. Shown turbine is WT 7, using SKR filter. Zoomed detail (top right).}
\label{fig:ecdf-Cantelli}
\end{figure}

For visualization and offline threshold selection, Fig.~\ref{fig:ecdf-Cantelli} shows the empirical CDF of \(J_j(k)\) computed from healthy data, together with (i) the distribution-free Cantelli CDF bound; (ii) a horizontal line at the per-test confidence level \(1-\alpha_{\text{single}}\); and (iii) a vertical marker at two times the Cantelli threshold. The Markov bound is much looser; its threshold \(2h_{\text{Markov}}\) lies far to the right and is not shown in the figure, but its numerical value appears in Table~\ref{tab:thresholds}. The example figure is drawn for a representative turbine using the SKR filter; the same procedure is applied identically to the unified solution implementation.

Tables~\ref{tab:calib_skr} and \ref{tab:calib_hinf} report, for each turbine and for each filter, the calibration statistics used in the distribution-free threshold selection approach. The empirical mean \(\hat\mu_J\) and standard deviation \(\hat\sigma_J\) are computed by pooling all windowed statistics \(J_j(k)\) obtained from healthy data. The effective multiplicity \(M_{\mathrm{eff}}\) is estimated on the same segment by forming, at fixed \(k\), the vector \(\{J_j(k)\}_{j=k-M+1}^k\), building the columnwise correlation matrix across \(j\), and applying the Li--Ji eigenvalue rule. The per-candidate level $\alpha_{\mathrm{single}}$ then follows from Sidák’s correction. All turbine-specific thresholds for both filters (Cantelli and Markov) are listed in Table~\ref{tab:thresholds}. For deployment, a single farm-wide threshold is used per implementation by taking the maximum \(h_{\text{Cantelli}}\) across turbines, as shown in the last row of Table~\ref{tab:thresholds}.

\begin{table}[htbp]
\caption{Calibration statistics and effective multiplicity (healthy data, SKR).}
\centering
\footnotesize
\setlength{\tabcolsep}{5pt}
\renewcommand{\arraystretch}{1.16}
\begin{tabular}{@{}ccccc@{}}
\hline
\textbf{Turbine} & \boldmath$\hat\mu_J$ & \boldmath$\hat\sigma_J$ & \boldmath$M_{\mathrm{eff}}$ & \boldmath$\alpha_{\mathrm{single}}$ \\
\hline
WT 1 & 368.463 & 336.656 & 1.62 & 0.000616 \\
WT 2 & 349.957 & 329.564 & 1.88 & 0.000532 \\
WT 3 & 358.811 & 360.405 & 1.25 & 0.000799 \\
WT 4 & 383.319 & 379.09 & 1.42 & 0.000706 \\
WT 5 & 371.883 & 389.119 & 1.36 & 0.000735 \\
WT 6 & 304.092 & 405.538 & 1.18 & 0.000845 \\
WT 7 & 326.828 & 367.031 & 1.23 & 0.000815 \\
\hline
\end{tabular}
\label{tab:calib_skr}
\end{table}

\begin{table}[htbp]
\caption{Calibration statistics and effective multiplicity (healthy data, $\mathcal H_i/\mathcal H_\infty$).}
\centering
\footnotesize
\setlength{\tabcolsep}{5pt}
\renewcommand{\arraystretch}{1.16}
\begin{tabular}{@{}ccccc@{}}
\hline
\textbf{Turbine} & \boldmath$\hat\mu_J$ & \boldmath$\hat\sigma_J$ & \boldmath$M_{\mathrm{eff}}$ & \boldmath$\alpha_{\mathrm{single}}$ \\
\hline
WT 1 & 230.740 & 213.560 & 1.41 & 0.000707 \\
WT 2 & 251.204 & 243.192 & 1.33 & 0.000751 \\
WT 3 & 231.400 & 247.800 & 1.16 & 0.000862 \\
WT 4 & 254.112 & 273.755 & 1.20 & 0.000836 \\
WT 5 & 206.231 & 218.316 & 1.28 & 0.000781 \\
WT 6 & 165.585 & 211.296 & 1.18 & 0.000851 \\
WT 7 & 180.423 & 186.355 & 1.28 & 0.000784 \\
\hline
\end{tabular}
\label{tab:calib_hinf}
\end{table}

\begin{table}[htbp]
\caption{Per-turbine thresholds on $g$ and adopted farm-wide value.}
\centering
\footnotesize
\setlength{\tabcolsep}{3pt}
\renewcommand{\arraystretch}{1.16}
\begin{tabular}{@{}ccccc@{}}
\hline
 & \multicolumn{2}{c}{\boldmath$\mathcal H_i/\mathcal H_\infty$} & \multicolumn{2}{c}{\textbf{SKR}} \\
\textbf{Turbine} & \boldmath$h_{\mathrm{Cantelli}}$ & \boldmath$h_{\mathrm{Markov}}$ & \boldmath$h_{\mathrm{Cantelli}}$ & \boldmath$h_{\mathrm{Markov}}$\\
\hline
WT 1 & 4129.15 & 163127.92 & 6965.62 & 299196.85\\
WT 2 & 4562.05 & 167322.21 & 7317.11 & 328889.90\\
WT 3 & 4333.04 & 134165.86 & 6551.76 & 224523.14\\
WT 4 & 4860.16 & 152050.34 & 7329.07 & 271496.17\\
WT 5 & 4008.06 & 132063.58 & 7357.86 & 252848.61\\
WT 6 & 3702.90 & 97293.19 & 7125.55 & 179986.85\\
WT 7 & 3417.37 & 115113.59 & 6590.36 & 200589.28\\
\textbf{Farm-wide} & 4860.16 & --- & 7357.86 & ---\\
\hline
\end{tabular}
\label{tab:thresholds}
\end{table}

The quantitative results in this section are restricted to the $\SI{13}{m/s}$ dataset. The other nine datasets span different wind conditions, but comparative detection statistics and residual envelopes are not reported here. In particular, the similarity of the two filters in this example cannot establish equivalent performance below rated operation or at other turbulence levels.

\begin{figure}
    \centering
    \includegraphics[width=0.48\textwidth]{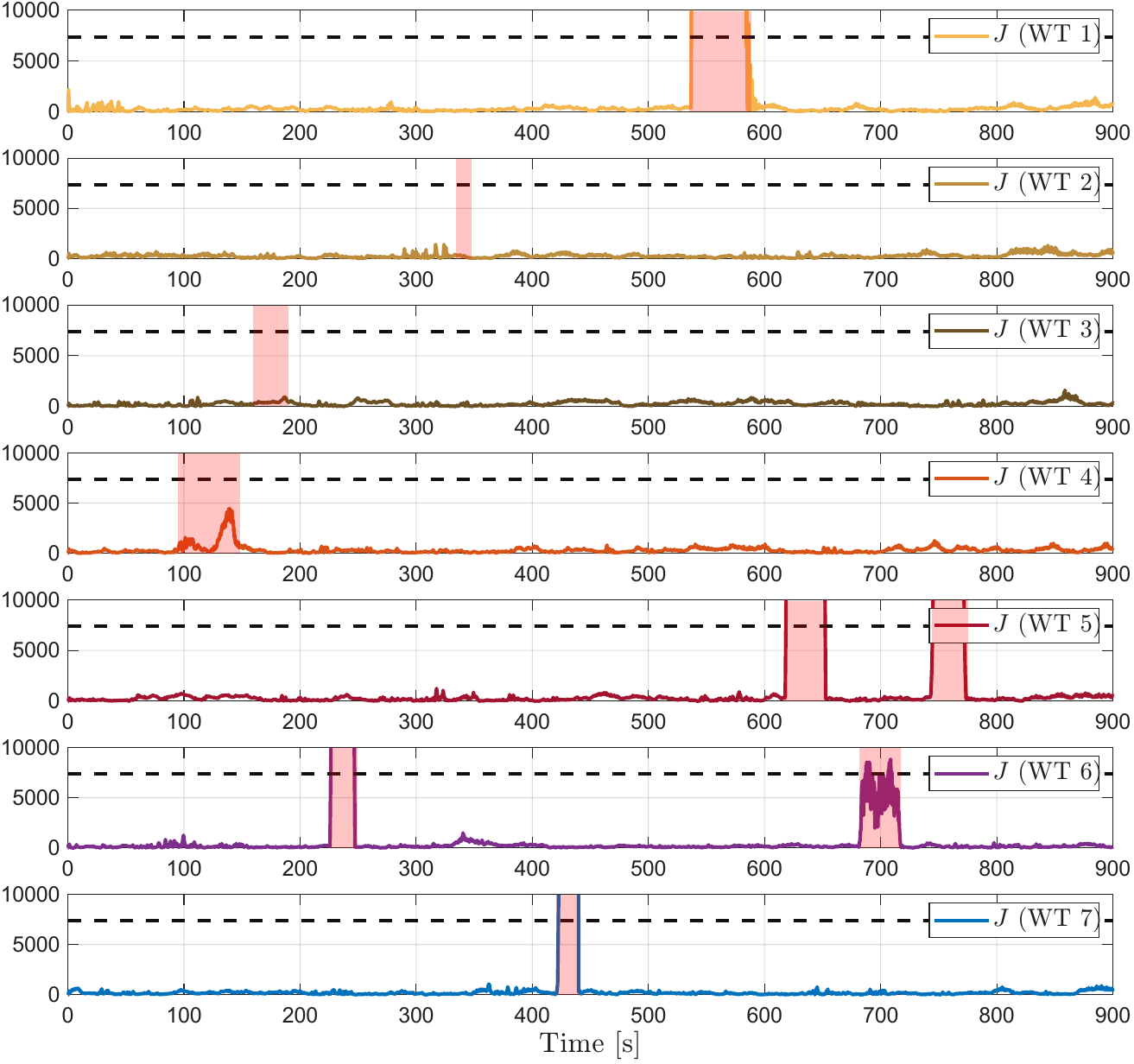}
    \caption{GLR decision functions and threshold $h_{\text{Cantelli}}=7357.86$ for the seven wind turbines of the wind farm, for the scenario with $\SI{13}{\meter/\second}$ mean wind speed and the SKR as the chosen filter. The faulty sections have been marked in red.}
    \label{fig:dec_fun_13ms_SKR}
\end{figure}

\begin{figure}
    \centering
    \includegraphics[width=0.48\textwidth]{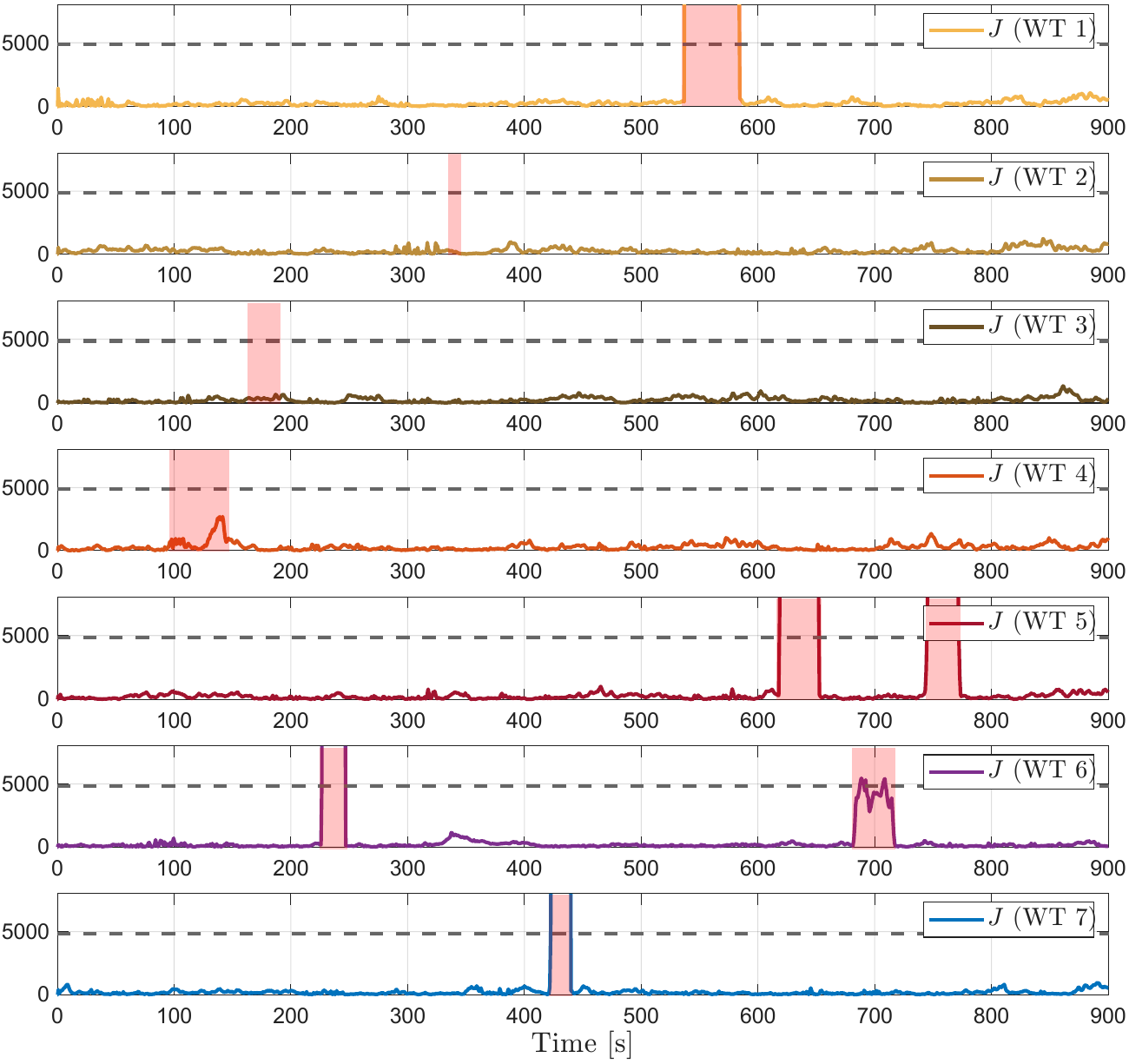}
    \caption{GLR decision functions and threshold $h_{\text{Cantelli}}=4860.16$ for the seven wind turbines of the wind farm, for the scenario with $\SI{13}{\meter/\second}$ mean wind speed and the unified solution. The faulty sections have been marked in red.}
    \label{fig:dec_fun_13ms_Hi_Hinf}
\end{figure}

Figs.~\ref{fig:dec_fun_13ms_SKR}--\ref{fig:dec_fun_13ms_Hi_Hinf} present the GLR decision functions for all seven turbines under the nine single-fault cases described in Sec.~\ref{sec:fault_scenarios}. Each fault is applied individually, and the corresponding farm-wide Cantelli threshold is used for each filter. Table~\ref{tab:fault_outcomes} summarizes the fault injections and the threshold-crossing outcomes for the unified solution only: six faults are detected and three remain undetected.

\begin{table*}[!t]
\caption{Injected single-fault cases at $\SI{13}{\meter/\second}$ mean wind speed and corresponding GLR responses for the unified solution. Detection delays are measured from fault onset.}
\label{tab:fault_outcomes}
\centering
\footnotesize
\setlength{\tabcolsep}{5pt}
\renewcommand{\arraystretch}{1.16}
\begin{tabular}{@{}c p{0.38\textwidth} c p{0.30\textwidth}@{}}
\hline
\textbf{WT} & \textbf{Injected fault} & \textbf{Interval (s)} & \textbf{Unified solution response (delay)} \\
\hline
4 & Pitch sensor gain, blade~3 ($+10\%$) & $[93.9,141.7]$ & Missed \\
3 & Generator speed sensor gain ($+5\%$) & $[150.9,186.7]$ & Missed \\
6 & Pitch sensor stuck & $[225.8,243.7]$ & Detected (0.7 s) \\
2 & Generator power sensor gain ($+10\%$) & $[328.8,341.7]$ & Missed \\
7 & Pitch actuator stuck & $[420.9,436.7]$ & Detected (1.6 s) \\
1 & Pitch sensor drift ($\kappa=4\times10^{-5}~\mathrm{deg/s}$) & $[537,579.2]$ & Detected (0.1 s) \\
5 & Pitch actuator offset ($3^{\circ}$) & $[617.9,652.3]$ & Detected (0.5 s) \\
6 & Generator torque actuator offset ($10^3~\mathrm{N\,m}$) & $[680.9,714.7]$ & Detected (6.3 s) \\
5 & Pitch actuator gain ($+10\%$) & $[742.9,770.7]$ & Detected (1.8 s) \\
\hline
\end{tabular}
\end{table*}

The comparison shows that the two observer designs propagate the same aerodynamic parameter uncertainty differently. As discussed in Sec.~\ref{subsec:bounds}, the SKR-based filter yields a smaller parametric uncertainty band for the rotor-speed residual, whereas the unified solution provides a whitened wind-speed residual and better decoupling of the actuator residuals. These differences are relevant when selecting residual channels and modelling their uncertainty. However, the GLR decision functions in Figs.~\ref{fig:dec_fun_13ms_SKR}--\ref{fig:dec_fun_13ms_Hi_Hinf} are very similar. This is due to the non-parametric covariance $R_{\mathrm{np}}$ added to $\Sigma_r(k)$ to account for unmodelled dynamics and prevent false alarms. Thus, for the model and fault scenarios considered, the differences in uncertainty propagation have limited impact on the final decision function. With higher-fidelity models requiring less covariance augmentation, these differences may become more important. The comparison does not establish that either design has superior detection performance.

The main objective is to propagate model uncertainty into the detection statistic to limit false alarms, rather than to maximize sensitivity to individual faults. In the reported scenario, the observed false-alarm rate is approximately $0.08\%$, while the missed-detection rate is $33\%$ (three out of nine faults). The latter remains a practical limitation. All three missed faults are sensor gain faults (Table~\ref{tab:fault_outcomes}). The generator-speed fault produces a deviation within the estimated residual variability, while the power-sensor fault is short and produces a relatively small deviation. For the pitch-sensor fault, a CUSUM detector using differential pitch residuals was reported in~\cite{gomez2025robust} to detect the fault at $t=106~\mathrm{s}$. This illustrates the benefit of residuals designed for specific subsystems.

The simplified model does not capture all turbine dynamics. The covariance term $R_{\mathrm{np}}$ accounts for part of this mismatch but also reduces the weight of affected residuals in the GLR. The conservative Cantelli threshold and the use of the largest turbine-specific threshold further reduce sensitivity. These choices help limit false alarms but may also mask weak faults. The excluded wind-speed and tower-acceleration residuals might provide indirect fault information. Further tests are needed to determine how much fault magnitude, residual sensitivity, uncertainty estimates, thresholding and channel exclusion each contribute to missed detections.

The aim is not to optimize detection for each fault, but to assess how model uncertainty can be accounted for when monitoring several subsystems with a common, simplified model. In practice, the method could serve as a general anomaly detector alongside dedicated subsystem detectors or data-driven methods. The reported missed-detection rate limits its use as a standalone safety-critical detector. Any improvement in sensitivity should be assessed together with the corresponding false-alarm rate.

Table~\ref{tab:metrics} summarizes the reported fault-detection and computation metrics for the $\SI{13}{\meter/\second}$ benchmark scenario. The detection outcomes (FAR, MDR and detection-delay range) refer to the detailed analysis of the unified solution. The reported average filter computation time applies to both designs.

\begin{table}[htbp]
\caption{Fault-detection outcomes and computation metrics.}
\centering
\footnotesize
\setlength{\tabcolsep}{5pt}
\renewcommand{\arraystretch}{1.16}
\begin{tabular}{@{}>{\raggedright\arraybackslash}p{0.60\columnwidth}>{\raggedright\arraybackslash}p{0.31\columnwidth}@{}}
\hline
\textbf{Metric} & \textbf{Reported value} \\
\hline
False alarm rate (FAR) & $\approx 0.08\%$ \\
Missed detection rate (MDR) & $33\%$ (3 of 9 faults) \\
Detection delay (reported range) & $0.1$--$\SI{6.3}{\second}$ \\
Benchmark runtime without FDI & $\SI{103.5}{\second}$ \\
Filter \& uncertainty propagation (per iteration; either design) & $\SI{2.3e-4}{\second}$ \\
GLR test (per iteration) & $\SI{0.0019}{\second}$ \\
Complete algorithm: mean per-iteration runtime ratio ($t_{\mathrm{sim+FDI}}/t_{\mathrm{sim}}$) & $\approx 1.00002058$ \\
\hline
\end{tabular}
\label{tab:metrics}
\end{table}

\section{Conclusion}\label{sec:conclusion}

This paper presented a comprehensive framework for robust, model-based fault detection in a floating offshore wind-farm benchmark, explicitly accounting for parametric uncertainty. Two observer-based residual generators, the orthogonal-projection SKR-based filter and the unified solution ($\mathcal{H}_i/\mathcal{H}_\infty$), were implemented and compared under identical conditions. Uncertainty propagation was derived analytically up to second order. The first-order covariance approximation was used in the GLR test, while second-order bounds were computed to assess the effect of nonlinear uncertainty propagation.

For the reported $\SI{13}{m/s}$ case, the detailed results for the unified solution show detection of the injected actuator faults with delays of $0.5$--$\SI{6.3}{s}$, whereas residuals tied to aerodynamics remain limited by model uncertainty. The unified solution whitens the wind–speed residual, which introduces a substantial estimation uncertainty; the SKR instead exposes the wind profile directly in that residual and therefore carries no estimation uncertainty on that channel, which decreases the uncertainty bands in the other SKR residuals due to coupling in the dynamics. In our setting this difference has little impact on the decision stage because faults are not injected in wind (or tower) sensors, so those residuals are not needed in the residual evaluation stage.

Despite the distinct filter designs, the two filters yield similar decision functions in the reported scenario. This is because, to prevent false alarms under unmodelled dynamics, the propagated residual parametric covariance needs to be augmented with small non–parametric terms; once this augmentation is applied, the effective uncertainty levels seen by the decision function become comparable across filters. Design differences may become more evident with higher-fidelity models and better-characterized parametric uncertainty.

The main contribution is the propagation of model uncertainty to support false-alarm control in the considered observer designs, rather than the optimization of detection sensitivity for individual faults. The reported scenario yields a low observed false-alarm rate ($0.08\%$), but a missed-detection rate of $33\%$, highlighting the limits of the simplified model and detection configuration, particularly for sensor gain faults. Future work should assess the detection--false-alarm trade-off across operating conditions, investigate higher-fidelity models and complementary subsystem-specific or data-driven detectors, and validate the approach with field measurements.



{\appendices

\section{Fault Detection Filters}
\label{appendixA}

\subsection{Notation}

$\mathcal{L}_2$ denotes the space of square–summable discrete–time signals (vector $\ell_2$ sequences) with norm
$\|x\|_2^2 = \sum_{k=0}^{\infty} \|x(k)\|_2^2$ and inner product
$\langle x, y \rangle_{\mathcal{L}_2} = \sum_{k=0}^{\infty} y(k)^{\ast} x(k)$.
$\mathcal{H}_2$ is the discrete–time Hardy space of stable, causal transfer matrices analytic in $|z|<1$ with finite $\mathcal{H}_2$ norm.
$\mathcal{R}\mathcal{H}_\infty$ denotes the set of proper, real–rational, stable transfer matrices with finite $\mathcal{H}_\infty$ norm
$\|G\|_\infty = \sup_{\omega \in [0,2\pi]} \bar\sigma\big(G(e^{j\omega})\big)$, where $\bar\sigma(\cdot)$ is the largest singular value.
For any stable transfer matrix $K(z)$, $\mathcal{L}_K$ is the multiplication operator $x \mapsto Kx$ acting on $\mathcal{L}_2$, and
$\mathcal{L}_K^{\sim}$ is its adjoint; equivalently, $\mathcal{L}_K^{\sim} = \mathcal{L}_{K^{\sim}}$ with paraconjugate
$K^{\sim}(z) := K^{\ast}(z^{-1})$.
Given a closed subspace $\mathcal{K} \subset \mathcal{L}_2$, $\mathcal{P}_{\mathcal{K}}$ denotes the orthogonal projector onto $\mathcal{K}$,
and $\mathcal{K}^{\perp} = \{x \in \mathcal{L}_2 : \langle x, y \rangle_{\mathcal{L}_2} = 0 \ \forall\, y \in \mathcal{K}\}$ is its orthogonal complement.
Throughout, $(\cdot)^{\ast}$ denotes conjugate transpose (for real–valued matrices it reduces to transpose).

\subsection{Coprime factorizations, and kernel/image representations}

We consider a discrete-time LTI plant
\begin{align}
x(k{+}1)&=Ax(k)+Bu(k),\label{eq:model_x_dot}\\
y(k)&=Cx(k)+Du(k).\label{eq:model_y}
\end{align}
Here $x\in\mathbb{R}^n$ (state), $u\in\mathbb{R}^p$ (input), $y\in\mathbb{R}^m$ (output). 
The system has minimal realization \((A,B,C,D)\) and proper real-rational transfer matrix $G(z)$ from the input to the output, $y(z)=G(z)u(z)$. Left/right coprime factorizations (LCF/RCF) over \(\mathcal{RH}_\infty\) are
\begin{equation}
G(z)=\hat M^{-1}(z)\hat N(z)=N(z)M^{-1}(z)\label{eqn:LCF_RCF}
\end{equation}
where, for stabilizing \(F,L\),
\begin{align}
\hat M(z)&=(A-LC,\,-L,\;WC,\;W)\label{eqn:M_hat}\\
\hat N(z)&=(A-LC,\;B-LD,\;WC,\;WD)\label{eqn:N_hat}\\
M(z)&=(A+BF,\;BV,\;F,\;V)\\
N(z)&=(A+BF,\;BV,\;C{+}DF,\;DV)
\end{align}

and the Bezout identity holds\cite{ding2021advanced}.

Define the stable kernel and image representations
\begin{equation}
    K_G(z)=\begin{bmatrix}-\hat N(z)&\hat M(z)\end{bmatrix},\qquad
    I_G(z)=\begin{bmatrix}M(z)\\N(z)\end{bmatrix},
\end{equation}
so that
\begin{equation}
    r(z)=K_G(z)\!\begin{bmatrix}u(z)\\y(z)\end{bmatrix},\qquad
\begin{bmatrix}u(z)\\y(z)\end{bmatrix}=I_G(z)v(z).\label{eqn:SKR_SIR}
\end{equation}
Any LTI residual generator can be parametrized as $r=R(z)K_G\left[u^T,y^T\right]^T$ with a stable dynamic post-filter $R(z)\in\mathcal{RH}_\infty$.

\subsection{Orthogonal projection (SKR-based) fault-detection filter}

The development below is based on Subsection~5.2, \textit{"A fault detection scheme with projection onto $\mathcal{L}_2$,"} in~\cite{ding2026alternative}.

The associated kernel and image subspaces to \eqref{eqn:SKR_SIR} in $\mathcal{H}_2$ are
\begin{align}
\mathcal K_G&=\Bigl\{\begin{bmatrix}u\\y\end{bmatrix}\in \mathcal{H}_2:K_G\!\begin{bmatrix}u\\y\end{bmatrix}=0\Bigr\},\\
\mathcal I_G&=\Bigl\{\begin{bmatrix}u\\y\end{bmatrix}\in \mathcal{H}_2:\begin{bmatrix}u\\y\end{bmatrix}=I_G v,\ v\in \mathcal{H}_2\Bigr\}.
\end{align}

Choosing gains $(L_0,W_0,F_0,V_0)$ yields normalized factors\cite{hoffmann1996normalized}
\begin{equation}
    K_GK_G^{\sim}=I,\qquad I_G^{\sim}I_G=I,\label{eqn:norm_conditions}
\end{equation}
where
\begin{align}
L_0&=(BD^\top+APC^\top)\,(I+DD^\top+CPC^\top)^{-1},\\
W_0&=U\,(I+DD^\top+CPC^\top)^{-1/2},\quad U U^\top=I,\\
F_0&=-\bigl(I+D^\top D + B^\top Q B\bigr)^{-1}\bigl(D^\top C + B^\top Q A\bigr),\\
V_0&=\bigl(I+D^\top D + B^\top Q B\bigr)^{-1/2}\Gamma,\quad \Gamma^\top\Gamma=I,
\end{align}
with $P>0$, $Q>0$ solving
\begin{align}
\begin{split}
P &= A P A^\top + B B^\top \\
  &\quad - (B D^\top + A P C^\top)\, R^{-1}\, (B D^\top + A P C^\top)^\top,
\end{split} \\
\begin{split}
Q &= A^\top Q A + C^\top C \\
  &\quad - (D^\top C + B^\top Q A)^\top\, \bar R^{-1}\, (D^\top C + B^\top Q A),
\end{split}
\end{align}
where $R = I + D D^\top + C P C^\top$ and $\bar R = I + D^\top D + B^\top Q B$.

From normalization, the following identity holds:
\begin{equation}
\mathcal{L}_{K_G}^{\sim}\mathcal{L}_{K_G}+\mathcal{L}_{I_G}\mathcal{L}_{I_G}^{\sim}=\mathcal{I}.
\end{equation}
Therefore $\mathcal{L}_{K_G}^{\sim}\mathcal{L}_{K_G}$ is the orthogonal projector onto $\mathcal K_G^\perp$ (the residual subspace), and\cite{georgiou1988computation}
\begin{align}
P_{\mathcal K_G}&=\mathcal{I}-\mathcal{L}_{K_G}^{\sim}\mathcal{L}_{K_G},\\
r_{\mathcal K_G}&=(\mathcal{I}-\mathcal{P}_{\mathcal K_G})\!\begin{bmatrix}u\\y\end{bmatrix}=\mathcal{L}_{K_G}^{\sim}\mathcal{L}_{K_G}\!\begin{bmatrix}u\\y\end{bmatrix},
\end{align}
with $\|r_{\mathcal K_G}\|_2=\biggl\|\mathcal{L}_{K_{G}}\!\begin{bmatrix}u\\y\end{bmatrix}\biggr\|_2$.

The residual can be implemented in observer form
\begin{align}
    \hat x(k{+}1)&=(A{-}L_0C)\hat x(k){+}(B{-}L_0D)u(k){+}L_0 y(k),\label{eqn:obs1}\\
    r(k)&=W_0\,\bigl(y(k)-C\hat x(k)-Du(k)\bigr),\label{eqn:obs2}
\end{align}
and for matched initials (or after transients)
\begin{equation}
    r(z)=\hat M(z)\,y(z)-\hat N(z)\,u(z)=K_G(z)\!\begin{bmatrix}u(z)\\y(z)\end{bmatrix}.
\end{equation}
Hence $\|r\|_2=\|r_{\mathcal K_G}\|_2$, i.e., the observer residual equals the $\mathcal{L}_2$-orthogonal projection residual onto $\mathcal K_G^\perp$.

\subsection{Unified solution ($\mathcal{H}_i/\mathcal{H}_\infty$)}

We consider a discrete-time LTI plant
\begin{align}
x(k{+}1)&=Ax(k)+Bu(k)+E_d\,d(k)+E_f\,f(k),\\
y(k)&=Cx(k)+Du(k)+F_d\,d(k)+F_f\,f(k).
\end{align}
Let the observer-based residual generator be
\begin{align}
\hat x(k{+}1)&=(A-LC)\hat x(k)+(B-LD)u(k)+Ly(k),\\
r(k)&=V\bigl(y(k)-C\hat x(k)-Du(k)\bigr).\label{eqn:obs_form}
\end{align}
Under disturbances \(d\) and faults \(f\), the observer-based residual generator induces the following mappings from $d$ and $f$ to the residual $r$, \(r(z)=V\hat N_d(z)\,d(z)+V\hat N_f(z)\,f(z)\), where transfer matrices $\hat N_d(z)$ and $\hat N_f(z)$ describe, respectively, the effects of disturbances and faults on the residual
\begin{align}
\hat N_d(z)&=F_d+C(zI-A+LC)^{-1}(E_d-LF_d),\\
\hat N_f(z)&=F_f+C(zI-A+LC)^{-1}(E_f-LF_f).
\end{align}
The robust design is then formulated as a trade-off between disturbance attenuation and fault sensitivity. Specifically, the optimal gains $\{L_{opt},V_{opt}\}$ are obtained by maximizing the singular values of $V\hat{N}_f(e^{j\theta})$ subject to an $\mathcal{H}_\infty$ bound on $V\hat{N}_d$
\begin{equation}
    \forall\theta\in [0, 2\pi], \{L_{opt},V_{opt}\} = \arg\sup_{L, V}\frac{\sigma_i\big(V\hat{N}_f(e^{j\theta})\big)}{||V\hat{N}_d||_{\infty}},
\end{equation}
with $i = 1, \dots, m$. This ratio is typically referred to as the $\mathcal{H}_i/\mathcal{H}_\infty$ performance index\cite{ding2008model}.

Assuming detectability and the surjectivity/regularity condition,
\[
\operatorname{rank}\begin{bmatrix}A-e^{j\theta} I&E_d\\ C&F_d\end{bmatrix}=n+m,\ \forall\theta\in[0,2\pi],
\]
the optimal observer gain and post-filter are
\begin{align}
    L_{opt} &= \left(APC^T+E_dF_d^T\right)V_{opt},\\
    V_{opt} &= \left(CPC^T+F_dF_d^T\right)^{-1/2},
\end{align}
where $P>0$ solves the Riccati equation
\begin{equation}
    APA^T-P+E_dE_d^T-L_{opt}\left(CPC^T+F_dF_d^T\right)L_{opt}^T=0.
\end{equation}
The unified solution solves the $\mathcal{H}_-/\mathcal{H}_{\infty}$ and $\mathcal{H}_{\infty}/\mathcal{H}_{\infty}$ design frameworks as special cases\cite{ding2008model}.

Let \(d(k)\sim\mathcal N(0,I)\) and interpret $w(k)=E_d\,d(k)$, $v(k)=F_d\,d(k)$, so that the process-noise covariance is \(Q=E_dE_d^\top\), the measurement-noise covariance is \(R=F_dF_d^\top\), and the (possibly nonzero) cross-covariance is \(S=E_dF_d^\top\). Then the Riccati equation and gain formula in the unified solution are exactly those of the steady-state (correlated-noise) Kalman filter. Moreover, the optimal post-filter is a measurement-noise whitener, \(V_{opt}=(F_dF_d^\top)^{-1/2}\), so the residual \(r=V_{opt}(y-C\hat x-Du)\) is the whitened innovation.



\section{Uncertainty Propagation for SKR-Based Orthogonal Projection Filter}
\label{appendixB}

We use a bookkeeping scalar \(\varepsilon\) and expand any perturbed quantity as
\(Z(\varepsilon)=Z_0+\varepsilon Z_1+\varepsilon^2 Z_2+\mathcal O(\varepsilon^3)\);
$Z_1$, $Z_2$ are the first- and second-order series coefficients.
Signal-level terms use lower case (e.g., $z_1$, $z_2$).
For generality, we allow perturbations in all system matrices \((A,B,C,D)\).

The covariance \(P\) is the stabilizing solution of the DARE
\begin{equation}
    \begin{split}
    P &= APA^\top + BB^\top\\
    &\quad- (BD^\top + APC^\top)\,R^{-1}\,(BD^\top + APC^\top)^\top,
    \end{split}
\end{equation}
where $R = I + DD^\top + CPC^\top$. The DARE can be written in the standard discrete-time algebraic Riccati form
\begin{equation}
    \begin{split}
    0 = &A_r^\top P A_r - P + Q_r - \big(A_r^\top P B_r + S_r\big)\\
    &\big(R_r + B_r^\top P B_r\big)^{-1}\big(B_r^\top P A_r + S_r^\top\big),
    \end{split}\label{eq:DARE-standard}
\end{equation}
with $A_r=A^\top$, $B_r=C^\top$, $Q_r=BB^\top$, $R_r=I+DD^\top$, $S_r=BD^\top$. Define $C_r=R_r+B_r^\top P B_r$, $D_r=A_r^\top P B_r+S_r$, $F=-C_r^{-1}D_r^\top$, $\Phi=A_r+B_rF$

and \(K=P\Phi\). Following Sun’s sensitivity analysis for the DARE \cite{sun1998sensitivity}, let the data be perturbed as
\(\tilde{A}_r=A_r+\Delta A_r\), \(\tilde{B}_r=B_r+\Delta B_r\), \(\tilde{R}_r=R_r+\Delta R_r\),
\(\tilde{S}_r=S_r+\Delta S_r\), \(\tilde{Q}_r=Q_r+\Delta Q_r\), and set
\begin{align*}
\delta R_r&=\Delta R_r+B_r^\top P\,\Delta B_r+\Delta B_r^\top P\,B_r+\Delta B_r^\top P\,\Delta B_r,\\
\delta S_r&=\Delta S_r+A_r^\top P\,\Delta B_r+\Delta A_r^\top P\,B_r+\Delta A_r^\top P\,\Delta B_r,
\end{align*}
Let $\tilde{C}_r=\tilde{R}_r+\tilde{B}_r^\top P \tilde{B}_r$, $\tilde{D}_r=\tilde{A}_r^\top P \tilde{B}_r+\tilde{S}_r$, $\tilde{F}=-\tilde{C}_r^{-1}\tilde{D}_r^\top$, $\tilde{\Phi}=\tilde{A}_r+\tilde{B}_r\tilde{F}$, $\Delta\Phi=\tilde{\Phi}-\Phi$.

If \(\tilde{P}\) solves the perturbed DARE, \(\Delta P:=\tilde{P}-P\) satisfies the perturbation equation
\begin{equation}
\Delta P - \Phi^\top \Delta P\,\Phi \;=\; E_1 + E_2 + h_1(\Delta P) + h_2(\Delta P),
\label{eqn:Delta_P}
\end{equation}
with
\begin{align}
E_1 &= K^\top \Delta A_r+\Delta A_r^\top K+K^\top \Delta B_r F\nonumber\\
&\quad +F^\top \Delta B_r^\top K+F^\top \Delta R_r F+\Delta S_r F\nonumber\\
&\quad+F^\top \Delta S_r^\top+\Delta Q_r,\\
E_2 &= \Delta A_r^\top P \Delta A_r+\Delta A_r^\top P \Delta B_r F\nonumber\\
&\quad +F^\top \Delta B_r^\top P \Delta A_r+F^\top \Delta B_r^\top P \Delta B_r F \nonumber\\
&\quad -\big(F^\top \delta R_r+\delta S_r\big)\big(I+C_r^{-1}\delta R_r\big)^{-1} C_r^{-1}\delta S_r^\top\nonumber\\
&\quad -\delta S_r C_r^{-1}\delta R_r\big(I+C_r^{-1}\delta R_r\big)^{-1} F,\\
h_1(\Delta P) &= \Phi^\top \Delta P\,\Delta\Phi+\Delta\Phi^\top \Delta P\,\Phi\nonumber\\
&\quad +\Delta\Phi^\top \Delta P\,\Delta\Phi,\\
h_2(\Delta P) &= -\,\tilde{\Phi}^\top \Delta P\,\tilde{B}_r\big(I+\tilde{C}_r^{-1}\tilde{B}_r^\top \Delta P\,\tilde{B}_r\big)^{-1}\nonumber\\
&\quad \tilde{C}_r^{-1}\tilde{B}_r^\top \Delta P\,\tilde{\Phi}.
\end{align}
Using the series convention, write \(P(\varepsilon)=P+\varepsilon P_1+\varepsilon^2 P_2+\mathcal O(\varepsilon^3)\).
Then \eqref{eqn:Delta_P} yields
\begin{align}
P_1 - \Phi^\top P_1 \Phi &= E_1,\label{eqn:P1}\\
P_2 - \Phi^\top P_2 \Phi &= E_2 + h_1(P_1) + h_2(P_1).\label{eqn:P2}
\end{align}


First- and second-order uncertainties are subsequently propagated through the filter gain
\begin{equation}
\begin{split}
    L = \mathcal A\mathcal B^{-1}, \quad
    &\mathcal A = BD^\top + APC^\top,\\
    &\mathcal B = I + DD^\top + CPC^\top,
\end{split}
\end{equation}
the post-filter
\begin{equation}
    W \;=\; \mathcal B^{-1/2},
\end{equation}
and the residual map $r(k) = W(y(k)-\hat y(k))$.

The induced expansions $\mathcal A(\varepsilon)$ and $\mathcal B(\varepsilon)$ are obtained by substituting $A(\varepsilon),B(\varepsilon),C(\varepsilon),D(\varepsilon),P(\varepsilon)$
into the definitions above and collecting powers of~$\varepsilon$,
\begin{align}
\mathcal A_0 &= BD^\top+APC^\top,\\
\mathcal B_0 &= I+DD^\top+CPC^\top,\\
\mathcal A_1 &= B_1D^\top + B D_1^\top \nonumber\\
               &\quad \;+\; A_1PC^\top + AP_1C^\top + AP C_1^\top,\\
\mathcal B_1 &= D_1D^\top + D D_1^\top \nonumber\\
               &\quad \;+\; C_1PC^\top + CP_1C^\top + CP C_1^\top ,\label{eq:ABfirst}\\
\mathcal A_2 &= B_2D^\top + B D_2^\top + B_1D_1^\top \nonumber\label{eq:A2}\\
             &\quad + A_2PC^\top + AP_2C^\top + APC_2^\top\nonumber\\
             &\quad + A_1P_1C^\top + A_1PC_1^\top + AP_1C_1^\top,\\
\mathcal B_2 &= D_2D^\top + D D_2^\top + D_1D_1^\top \nonumber\label{eq:ABsecond}\\
             &\quad + C_2PC^\top + CP_2C^\top + CPC_2^\top\nonumber\\
              &\quad + C_1P_1C^\top + C_1PC_1^\top + CP_1C_1^\top.
\end{align}

Second-order expansion of $\mathcal B(\varepsilon)^{-1}$; let $G(\varepsilon)=\mathcal B(\varepsilon)^{-1}=\sum_{k=0}^2 \varepsilon^k G_k+\mathcal O(\varepsilon^3)$. From $\mathcal B(\varepsilon)G(\varepsilon)=I$,
\begin{align}
G_0&=\mathcal B_0^{-1},\\
G_1&=-\,\mathcal B_0^{-1}\mathcal B_1\mathcal B_0^{-1},\\
G_2&=\mathcal B_0^{-1}\mathcal B_1\mathcal B_0^{-1}\mathcal B_1\mathcal B_0^{-1}-\mathcal B_0^{-1}\mathcal B_2\mathcal B_0^{-1}.
\label{eq:G012}
\end{align}

Second-order expansion of $L(\varepsilon)=\mathcal A(\varepsilon)\,G(\varepsilon)$;
with $L(\varepsilon)=\sum_{k=0}^2\varepsilon^k L_k+\mathcal O(\varepsilon^3)$ and $L_k=\sum_{i+j=k}\mathcal A_iG_j$,
\begin{align}
L_0 &= \mathcal A_0\,\mathcal B_0^{-1},\label{eqn:L0}\\
L_1 &= \mathcal A_1\,\mathcal B_0^{-1} \;-\; \mathcal A_0\,\mathcal B_0^{-1}\mathcal B_1\mathcal B_0^{-1}, \label{eqn:L1}\\
L_2 &= \mathcal A_2\,\mathcal B_0^{-1} \;-\; \mathcal A_1\,\mathcal B_0^{-1}\mathcal B_1\mathcal B_0^{-1}\nonumber\\
&\quad + \mathcal A_0\Big(\mathcal B_0^{-1}\mathcal B_1\mathcal B_0^{-1}\mathcal B_1\mathcal B_0^{-1}
      - \mathcal B_0^{-1}\mathcal B_2\mathcal B_0^{-1}\Big).\label{eqn:L2}
\end{align}

Let $f(x)=x^{-1/2}$. With $W(\varepsilon)=f(\mathcal B(\varepsilon))=\sum_{k=0}^2\varepsilon^k W_k+\mathcal O(\varepsilon^3)$, applying the Fr\'echet–Taylor expansion of \(f\) at a SPD matrix \(\mathcal B_0\) and collecting powers of $\varepsilon$ yields
\begin{align}
W_0 &= \mathcal B_0^{-1/2},\label{eqn:W0}\\
W_1 &= \mathcal L_f[\mathcal B_0](\mathcal B_1),\\
W_2 &= \mathcal L_f[\mathcal B_0](\mathcal B_2) + \tfrac12\,\mathcal L_f^{(2)}[\mathcal B_0](\mathcal B_1,\mathcal B_1),
\end{align}
where \(\mathcal L_f\) and \(\mathcal L_f^{(2)}\) are the first and second Fr\'echet derivatives of the matrix function. By the Dalecki\u{\i}–Kre\u{\i}n theorem\cite{daleckii2002stability}, if \(\mathcal B_0 = U \Lambda U^\top\) with \(\Lambda=\mathrm{diag}(\lambda_1,\dots,\lambda_n)\) and \(U\) orthogonal, then in the eigenbasis, writing $E_1=U^\top \mathcal B_1 U$ and $E_2=U^\top \mathcal B_2 U$,
\begin{align}
    (W_1)_\sharp &= F_1\circ E_1,\\
    (W_2)_\sharp &= F_1\circ E_2 + \tfrac12\sum_{k=1}^n F_2^{(k)}\circ\big(E_1(:,k)\,E_1(k,:)\big),
\end{align}
where $(F_1)_{ij}=f^{[1]}(\lambda_i,\lambda_j)$ and $(F_2^{(k)})_{ij}=f^{[2]}(\lambda_i,\lambda_k,\lambda_j)$ are the first and second divided differences of $f$; use the diagonal limits $f^{[1]}(\lambda,\lambda)=f'(\lambda)=-\tfrac12\lambda^{-3/2}$ and $f^{[2]}(\lambda,\lambda,\lambda)=\tfrac12 f''(\lambda)=\tfrac{3}{8}\lambda^{-5/2}$ for coincident/near-coincident eigenvalues. Transforming back,
\begin{align}
W_1 &{=} U\,(F_1\circ E_1)U^\top,\label{eqn:W1}\\
W_2 &{=} U\!\left(\hspace{-1mm} F_1{\circ} E_2 {+} \tfrac12\sum_k F_2^{(k)}{\circ}\big(E_1(:,k)E_1(k,:)\big)\hspace{-1mm}\right)\!U^\top\label{eqn:W2}.
\end{align}

Consider the discrete-time observer
\begin{align}
\hat x(k{+}1) &= (A{-}LC)\hat x(k)
 + (B{-}LD)u(k)
 + L\,y(k), \label{eq:obs}\\
\hat y(k) &= Cx(k) + Du(k), \label{eq:yhat}
\end{align}
with measured $y(k),u(k)$. Define the innovation and residual $e(k)=y(k)-\hat y(k)$, $r(k)=W e(k)$.
Let $\hat x_1$, $\hat x_2$, $\hat y_1$, $\hat y_2$ denote the first-
and second-order perturbations. Collecting orders in \eqref{eq:obs} yields
\begin{align}
\hat x_1(k{+}1) &=
(A-LC)\,\hat x_1(k)\nonumber \\
&\quad + (A_1 - L_1 C - L C_1)\,\hat x(k)\nonumber \\
&\quad + (B_1 - L_1 D - L D_1)\,u(k)\nonumber \\
&\quad + L_1\,y(k),
\label{eq:dx1}\\
\hat x_2(k{+}1) &=
(A-LC)\,\hat x_2(k)\nonumber \\
&\quad + (A_1 - L_1 C - L C_1)\,\hat x_1(k)\nonumber \\
&\quad + (A_2 - L_2 C - L_1 C_1 - L C_2)\,\hat x(k)\nonumber \\
&\quad + (B_2 - L_2 D - L_1 D_1 - L D_2)\,u(k)\nonumber \\
&\quad + L_2\,y(k).
\label{eq:dx2}
\end{align}
From \eqref{eq:yhat},
\begin{align}
\hat y_1(k) &= C\, \hat x_1(k)
+ C_1\,\hat x(k)
+ D_1\,u(k),
\label{eq:dy1}\\[3pt]
\hat y_2(k) &=
C\,\hat x_2(k)
+ C_1\,\hat x_1(k)\notag\\
&\quad + C_2\,\hat x(k)
+ D_2\,u(k).
\label{eq:dy2}
\end{align}
With $e(k)(\varepsilon)=y(k)-\hat y(k)(\varepsilon)=e(k)-\varepsilon\hat y_1(k)-\varepsilon^2\hat y_2(k)+\mathcal O(\varepsilon^3)$ and $r(k)(\varepsilon)=W(\varepsilon)e(k)(\varepsilon)$, the residual perturbations are
\begin{align}
r_1(k) &= W_1e(k) - W\hat y_1(k),
\label{eq:dr1}\\[3pt]
r_2(k) &= W_2e(k) - W\hat y_2(k) - W_1\hat y_1(k).
\label{eq:dr2}
\end{align}

}


\bibliographystyle{IEEEtran} 
\bibliography{Bibliography} 

\vspace{11pt}

\vspace{-33pt}
\begin{IEEEbiography}[{\includegraphics[width=1in,height=1.25in,clip,keepaspectratio]{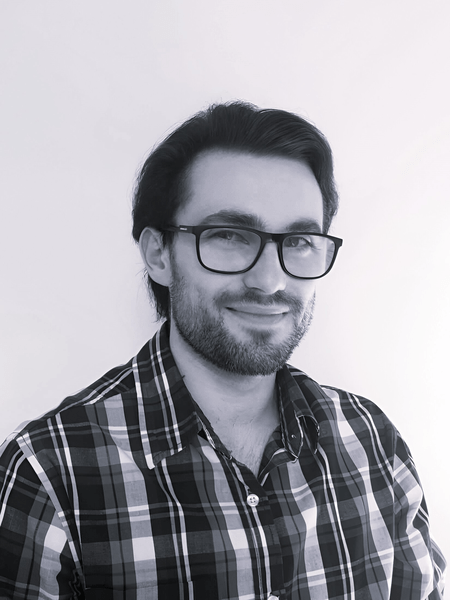}}]{Álvaro Martín Gómez}
received the M.Sc. in Control and Automation from Aalborg University, Aalborg, Denmark in 2022. He is currently pursuing a Ph.D. within the Department of Electronic Systems at the same university, and he is part of the Learning and Decisions Lab research group. His research focuses on the integration of physical knowledge of dynamical systems with data-driven techniques to overcome the limitations inherent in each approach individually. He seeks to develop new techniques for the accurate and efficient detection of anomalies and cyber-attacks in wind farms.
\end{IEEEbiography}

\begin{IEEEbiography}[{\includegraphics[width=1in,height=1.25in,clip,keepaspectratio]{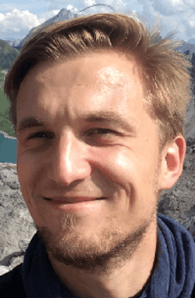}}]{Szymon Greś}
obtained his Ph.D. degree in Structural Dynamics from Aalborg University in 2019. After postdoctoral stays at Inria (French National Institute for Research in Computer Science and Automatic Control) in Rennes, France (2019) and ETH in Z{\"u}rich (2021), he joined Automation and Control section at Aalborg University as a postdoctoral researcher (2024).  His research interests are in data-driven methods for identification, fault diagnosis and uncertainty quantification in dynamical systems. 

\end{IEEEbiography}

\begin{IEEEbiography}[{\includegraphics[width=1in,height=1.25in,clip,keepaspectratio]{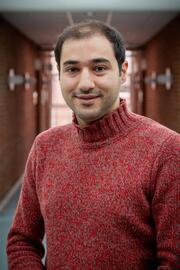}}]{Sina Hassani}
received his Ph.D. degree in Control Theory and its Applications to Power Systems from Aalborg University in 2025. He is currently a Research Assistant at Aalborg University and a member of the Learning and Decision Lab research group. His research interests are in control theory and its applications to energy systems, as well as cyber-attack diagnosis and mitigation techniques.
\end{IEEEbiography}

\begin{IEEEbiography}[{\includegraphics[width=1in,height=1.25in,clip,keepaspectratio]{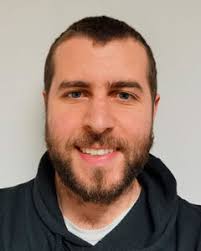}}]{Yerai Peña-Sánchez}
received the Diploma in Renewable Energies Engineering from the University of the Basque Country (EHU), Eibar, Spain, in 2016. He completed his PhD in 2020 at the Centre for Ocean Energy Research (COER), Maynooth University, where his research focused on the estimation and forecasting of wave excitation forces. Following his PhD, he carried out postdoctoral research at COER for one year, during which he developed two experimental test benches. He subsequently held a postdoctoral position in the Department of Mathematics at EHU, working on fault detection and isolation (FDI) and fault-tolerant control (FTC) strategies for offshore wind farms. He is currently a postdoctoral researcher at Mondragon University, focusing on offshore energy applications.
\end{IEEEbiography}

\begin{IEEEbiography}[{\includegraphics[width=1in,height=1.25in,clip,keepaspectratio]{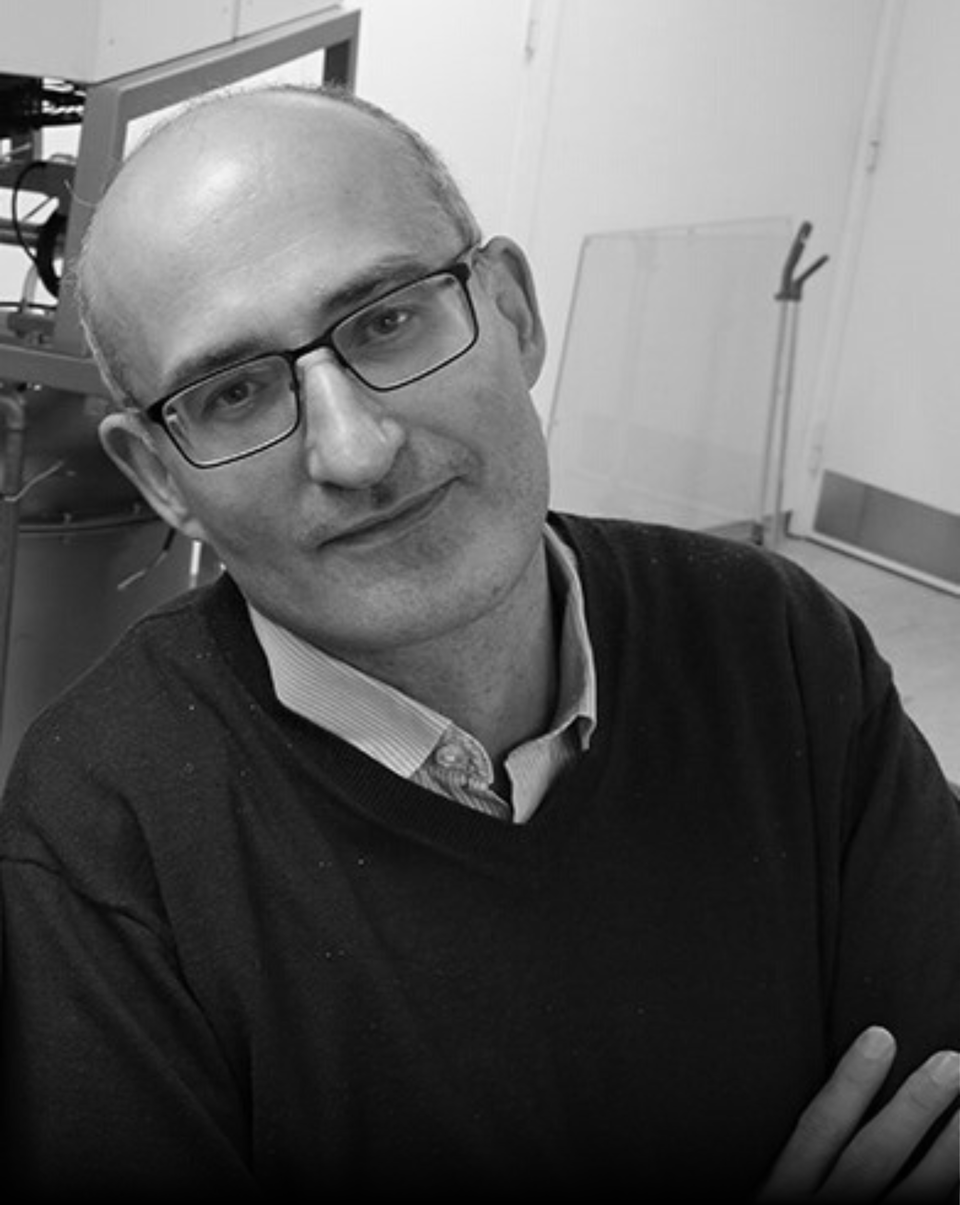}}]{Rafał Wiśniewski}
obtained his first Ph.D. degree in 1997 in the Department of Control Engineering, and his second Ph.D. degree in 2005 in the Department of Mathematical Sciences, both at Aalborg University, Aalborg, Denmark. He is a professor in the Department of Electronic Systems at the same university, in the section of Automation and Control, and the Head of Doctoral Program in Electrical and Electronic Engineering. His research objectives are to establish mathematical formalism for analysis of cyber-physical systems, and to develop numerical methods for solving fundamental control problems for non-linear systems: stability, robustness, and safety. He has over 50 peer reviewed international journal papers, and over 100 peer reviewed conference papers. He received the Aalborg City Springtime Award in 1999 for his contributions of the Satellite Team at Aalborg University to advance the status of society and the recognition of the city. He secured major grants with him as a principal investigator, such as CodeMe, EDGE, Smart $\&$ Cool, Wind Turbine Control Program and Hybrid Control.
\end{IEEEbiography}

\begin{IEEEbiography}[{\includegraphics[width=1in,height=1.25in,clip,keepaspectratio]{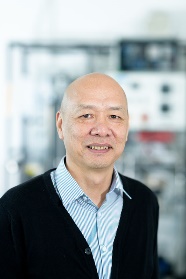}}]{Steven X. Ding}
received the Dr.-Ing. degree in
electrical engineering from the Gerhard-Mercator
University of Duisburg, Duisburg, Germany, in 1992.
From 1992 to 1994, he was a Research and
Development Engineer with Rheinmetall GmbH,
Düsseldorf, Germany. From 1995 to 2001, he was
a Full Professor of control engineering with
Lausitz University of Applied Sciences, Senftenberg, Germany, where he was the Vice President
from 1998 to 2000. Since 2001, he has been a
Full Professor of control engineering and the Head
of the Institute for Automatic Control and Complex Systems, University
of Duisburg-Essen, Duisburg. His research interests include model-based
and data-driven fault diagnosis, fault-tolerant systems, and their applications
in industry with a focus on automotive systems, chemical processes, and
renewable energy systems.
\end{IEEEbiography}

\end{document}